\documentclass[11pt, a4paper, logo, copyright, nonumbering]{cls/Sci}
\usepackage[numbers, sort&compress, square]{natbib}
\usepackage{dblfloatfix}
\usepackage{ulem}
\usepackage{caption}
\usepackage{dramatist}
\usepackage{wrapfig}
\usepackage{xspace}
\usepackage{pifont} %
\usepackage{multirow}

\usepackage{xltabular}
\usepackage{longtable}
\usepackage{hyperref}
\usepackage{algorithm}
\usepackage{algpseudocode}
\usepackage{amsmath}
\usepackage{amssymb}
\usepackage{amsfonts}
\usepackage{amsmath}
\usepackage{amssymb}
\usepackage{lineno}
\usepackage{multirow}
\usepackage{adjustbox}
\usepackage{cleveref}
\usepackage[bottom]{footmisc}
\usepackage{CJKutf8}
\usepackage{setspace}
\usepackage{bbding}
\usepackage{pifont}
\usepackage{dsfont}
\usepackage{array} %
\usepackage{tabularx} %
\usepackage{xcolor} %
\usepackage{tabularx}
\usepackage{booktabs}
\usepackage{tcolorbox}
\tcbuselibrary{breakable}
\usepackage{lipsum}  %
\usepackage{multicol} %
\definecolor{chatgray}{RGB}{245,245,245}
\definecolor{chatborder}{RGB}{200,200,200}
\usepackage{float}  %

\usepackage{listings}

\newtcolorbox{promptbox}[1][]{
  breakable,                
  colback=chatgray,         
  colframe=chatborder,      
  coltitle=black,           
  fonttitle=\bfseries,      
  title={#1},               
  arc=3mm,                  
  boxrule=0.5pt,           
  left=2mm, right=2mm, top=2mm, bottom=2mm, 
  fontupper=\small\ttfamily 
}

\makeatletter
\def\@BTrule[#1]{%
  \ifx\longtable\undefined
    \let\@BTswitch\@BTnormal
  \else\ifx\hline\LT@hline
    \nobreak
    \let\@BTswitch\@BLTrule
  \else
     \let\@BTswitch\@BTnormal
  \fi\fi
  \global\@thisrulewidth=#1\relax
  \ifnum\@thisruleclass=\tw@\vskip\@aboverulesep\else
  \ifnum\@lastruleclass=\z@\vskip\@aboverulesep\else
  \ifnum\@lastruleclass=\@ne\vskip\doublerulesep\fi\fi\fi
  \@BTswitch}
\makeatother

\addto\extrasenglish{
}

 {\begin{list}{}%
         {\setlength{\leftmargin}{#1}}%
         \item[]%
 }
 {\end{list}}
 
\reportnumber{001} %

\newcolumntype{Y}{>{\centering\arraybackslash}X}
\title{\centering SWE-Dev: Evaluating and Training Autonomous End-to-End Feature-Driven Software Development}

\vspace{-3em}
\author{
    Yaxin Du\textsuperscript{\rm 1, $\dagger$},
    Yuzhu Cai\textsuperscript{\rm 1,2 $\dagger$}, 
    Yifan Zhou\textsuperscript{\rm 1}, 
    Cheng Wang\textsuperscript{\rm 1},
    Yu Qian\textsuperscript{\rm 1},
    
    Xianghe Pang\textsuperscript{\rm 1},
    Qian Liu,
    Yue Hu\textsuperscript{\rm 2},
    Siheng Chen\textsuperscript{\rm 1, $\ast$},
    \\
    \textsuperscript{\rm 1} Shanghai Jiao Tong University, 
    \textsuperscript{\rm 2} Beijing University of Aeronautics and Astronautics, \textsuperscript{\rm 3} The Chinese University of Hong Kong, \textsuperscript{\rm 4} University of Michigan \\
    
}

\begin{document}
\begingroup
    \renewcommand{\thefootnote}{\fnsymbol{footnote}} 
    
    \footnotetext[2]{Equal contribution. * Corresponding author: \texttt{sihengc@sjtu.edu.cn}} 

\endgroup

\begin{abstract}
Large Language Models (LLMs) have shown strong capability in diverse software engineering tasks. However, feature-driven development, a highly prevalent real-world task that involves developing new functionalities for large, existing codebases, remains underexplored. We therefore introduce SWE-Dev, the first large-scale dataset (with 14,000 training and 500 test samples) designed to evaluate and train autonomous coding systems on real-world end-to-end feature-driven software development tasks. To ensure verifiable and diverse training, SWE-Dev uniquely provides all instances with a runnable environment and its developer-authored executable unit tests. 
This collection not only provides high-quality data for Supervised Fine-Tuning (SFT), but also enables Reinforcement Learning (RL) by delivering accurate reward signals from executable unit tests. We evaluated SWE-Dev across 17 base LLMs, 10 reasoning-focused LLMs, 10 multi-agent systems, and 8 tool-augmented LLM agents. Results show substantial headroom: the best single-turn model reaches only 22.51\% Pass@1 on the hard split, while OpenHands agents improve to 56.44\% but still leave many tasks unsolved. Code is available here \href{https://github.com/DorothyDUUU/SWE-Dev}{https://github.com/DorothyDUUU/SWE-Dev}.
\end{abstract}

\maketitle

\newpage

\begin{spacing}{0.9}
\end{spacing}

\newpage

\section{Introduction}

\begin{figure*}[b]
    \centering
    \includegraphics[width=\textwidth]{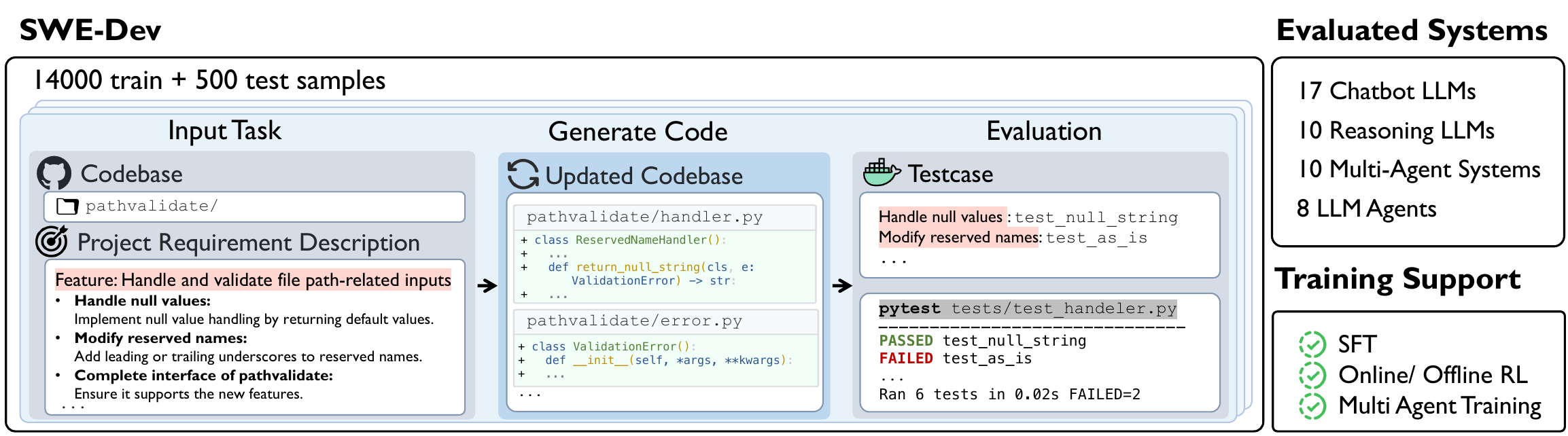}
    \caption{Overview of SWE-Dev, a software development dataset providing feature development tasks with feature description and codebase as input and test cases for evaluation. It is uniquely grounded in real-world repositories and paired with executable test suites, enabling reliable, functionally verifiable supervision. SWE-Dev is evaluated on 45 autonomous coding systems and supports advanced training paradigms like SFT, RL, and multi-agent training.}
    \label{fig:main}
\end{figure*}

Large Language Models (LLMs) are rapidly transforming autonomous programming, with capabilities extending from generating isolated code snippets to complex interactions within entire repositories ~\cite{github_copilot, cursor_ide}. As LLMs increasingly engage at this repository scale, rigorously evaluating their proficiency in handling complex coding systems becomes paramount for guiding their advancement. Current prominent benchmarks, while valuable, still struggle to judge how well LLMs perform in realistic, end-to-end development settings (Table~\ref{tab:comp_tab}). For example, SWE-Bench~\cite{swebench} measures only localized bug fixes described by GitHub issues, and RepoBench~\cite{repobench} evaluates the completion of a few unrelated functions within a repository.

In contrast, real-world software development is dominated by feature-driven development, where developers design, implement, and integrate new functionality into large, existing codebases. Prior empirical studies show that feature development accounts for a substantial fraction of daily engineering work ~\cite{harness2025getting_started_fdd, thum2014featureide}. Traditional feature development~\cite{coad1999java, palmer2002practical} typically involves intermediate processes such as developing an overall model or building feature lists. For coding systems, the  critical way towards achieving more comprehensive and genuinely autonomous programming capabilities is on the final implementation of features. 
This end-to-end process, from interpreting requirements within large, existing codebases to generating functionally correct and integrated code, is what we refer to as end-to-end feature-driven software development.

Recognizing the central role of feature development and the limitations of current evaluation benchmarks, we introduce an feature-driven \textbf{S}oft\textbf{W}ar\textbf{E} \textbf{Dev}elopment dataset, \textbf{SWE-Dev}. 
which is the first large-scale dataset designed to evaluate and train autonomous AI systems on real-world end-to-end feature-driven software development tasks. It comprises 14,000 training and 500 test instances derived from over 1,000 open-source projects, and is distinguished by three key characteristics: 
\ding{182} \textbf{Realistic scale and complexity}: SWE-Dev requires substantial code modifications (avg. 190 lines of code across 3 files), challenging models with cross-file dependencies, long contexts, and feature-level implementation scope.
\ding{183} \textbf{Robust and grounded evaluation}: Each sample is grounded in a real open-source repository, specified by a project requirement description (PRD), and evaluated using executable test cases to verify satisfaction of the intended implementation, ensuring reliable assessment and supervision.
\ding{184} \textbf{Verifiable training with executable test suites}: All 14,000 training instances include runnable environments and executable unit tests, providing execution-based feedback that supports supervised fine-tuning (SFT), reinforcement learning (RL) with accurate rewards, and multi-agent system (MAS) training.

We validate SWE-Dev from both evaluation and training perspectives. \textbf{For evaluation}, we test (i) base LLMs (instruction-tuned and reasoning-focused), (ii) tool-augmented LLM agents implemented via OpenHands, and (iii) multi-agent LLM systems (MAS). The results show that feature development remains challenging: under direct single-turn generation, the best base model Claude-3.7-Sonnet achieves only 49.47\% and 22.51\% Pass@1 on the \textit{easy} and \textit{hard} split, with most models far below this level, indicating substantial headroom. In contrast, OpenHands agents substantially improve performance, reaching 56.44\% on the hard split, but still leaving nearly half of tasks unsolved. Within our MAS suite, gains over a single baseline are modest (11.09\% $\rightarrow$ 20.03\% on hard), and simple refinement-based strategies can be more efficient than workflow-heavy designs. \textbf{For training}, SWE-Dev provides runnable tests and executable environments for every instance, enabling SFT and RL with execution feedback; we observe clear gains from test-supervised training, with a fine-tuned 7B model reaching performance comparable to GPT-4o on the hard split.

Our contributions are as follows:

\ding{182} We introduce \textbf{SWE-Dev}, the first real-world dataset for autonomous end-to-end feature-driven software development. The dataset includes both training and test splits, each with runnable environments and test cases, enabling a wide range of evaluation and training.

\ding{183} We conduct a comprehensive evaluation on SWE-Dev across \textbf{LLMs} (chatbot and reasoning-focused), tool-augmented \textbf{LLM agents} (via OpenHands), and \textbf{multi-agent LLM systems}, yielding insights into their capabilities and failure modes on feature development.

\ding{184} We demonstrate SWE-Dev \textbf{enabling and validating diverse training paradigms} (SFT, RL, and MAS training), establishing its utility for advancing training adaptation.

\begin{table}[t]
\centering
\caption{Comparison of SWE-Dev with existing repository-level benchmarks. We compare task type, the average number of golden lines of code, whether using the real-world executable tests are available, and the number of executable environments.}
\scalebox{0.85}{
\begin{tabular}{llccc}
\toprule
 & \textbf{Task type} & \textbf{\begin{tabular}[c]{@{}l@{}}Lines of \\ Code\end{tabular}} & \textbf{\begin{tabular}[c]{@{}l@{}}Real\\ Tests\end{tabular}} & \textbf{\begin{tabular}[c]{@{}l@{}}\#Exec \\ Env\end{tabular}} \\ \midrule
\textbf{ComplexCodeEval} & Func Comp & 38.2 & \color[HTML]{FE0000}\ding{55} & 0 \\
\textbf{CoderEval} & Func Comp & 20.6 & \color[HTML]{32CB00}\CheckmarkBold & 43 \\
\textbf{M2rc-Eval} & Line Comp & 1.0 & \color[HTML]{FE0000}\ding{55} & 0 \\
\textbf{CrossCodeEval} & Line Comp & 1.0 & \color[HTML]{FE0000}\ding{55} & 0 \\
\textbf{SWE-Bench}(Train) & Issue Solve & 32.8 & \color[HTML]{FE0000}\ding{55} & 0 \\
\textbf{R2E-Gym} & Issue Solve & 5.7 & \color[HTML]{FE0000}\ding{55} & 13 \\
\textbf{SWE-Smith} & Bug Fix & 9.6 & \color[HTML]{32CB00}\CheckmarkBold & 28 \\
\textbf{SWE-Synth} & Bug Fix & 5.7 & \color[HTML]{32CB00}\CheckmarkBold & 7 \\ \midrule
\rowcolor[HTML]{DAE8FC}
\textbf{SWE-Dev} & Feature Dev & 190.0 & \color[HTML]{32CB00}\CheckmarkBold & 1061 \\ \bottomrule
\end{tabular}}
\vskip -0.1in
\label{tab:comp_tab}
\end{table}

\section{Related Work}

\subsection{Coding benchmarks}
Early benchmarks such as HumanEval~\cite{humaneval}, MBPP~\cite{mbpp}, APPS~\cite{apps}, and CodeContests~\cite{codecontest} mainly target isolated, function-level problems with short contexts and well-specified I/O, which is useful for measuring basic correctness but far from real software development.
To better reflect repository-scale settings, a range of repository-level benchmarks have been proposed (Table~\ref{tab:comp_tab}), spanning function completion (e.g., ComplexCodeEval~\cite{complexcodeeval}, CoderEval~\cite{codereval}), line-level completion/understanding (e.g., M2RC-Eval~\cite{m2rc-eval}, CrossCodeEval~\cite{crosscodeeval}), and issue solving or bug fixing (e.g., SWE-Bench~\cite{swebench}, R2E-Gym~\cite{r2e-gym}, SWE-Smith~\cite{swe-smith}, SWE-Synth~\cite{swe-synth}). Despite this progress, two limitations remain common: (1) the scope of required code changes is still small (often \textless 40 LOC, e.g., 32.8 LOC in SWE-Bench and 38.2 LOC in ComplexCodeEval), which inadequately captures feature-scale development; and (2) evaluation is often weak or inconsistent, with many benchmarks lacking real executable tests or runnable environments, forcing reliance on proxy metrics that can miss functional correctness~\cite{starcoder, huang2022execution}. SWE-Dev directly addresses these gaps by providing large-scale, repository-level \emph{feature development} tasks with substantial code modifications and \textbf{real-world executable tests} in \textbf{runnable environments} for every instance, substantially improving realism and enabling training with execution feedback.

\subsection{Code LLMs Training}
Training LLMs for coding tasks typically involves three stages: pre-training, supervised fine-tuning (SFT), and reinforcement learning (RL). Pre-trained models such as StarCoder~\cite{starcoder} and Phi~\cite{abdin2024phi4technicalreport} leverage massive code corpora to learn syntax and general programming patterns. To improve instruction following and task completion, many works adopt SFT. Code Alpaca~\cite{codealpaca} employs self-instruct generation, WizardCoder~\cite{wizardcoder} leverages Evol-Instruct~\cite{wizardlm} to synthesize more complex instructions. However, SFT fundamentally lacks exploration: it teaches models to imitate ground-truth outputs rather than to reason or build autonomously~\cite{intructgpt}.
Beyond SFT, RL frameworks such as CodeRL~\cite{coderl} utilize test-case-based feedback to optimize model behavior. Early RL~\cite{rlhf} for code mostly targeted short, function-level problems with immediate execution rewards, but recent work has scaled RL to more realistic SWE settings: SWE-RL~\cite{swe-rl} applies policy-gradient updates in an agentless scaffold on SWE-Bench, while DeepSWE~\cite{deepswe} and SkyRL-v0~\cite{skyrl} (see also~\cite{golubev2025training}) train interactive, multi-turn SWE agents over full repositories and tools. In parallel, SWE-Gym~\cite{swe_gym} explores repository-scale training via multi-agent systems; however, due to the lack of an executable training split in SWE-Bench, it constructs a separate dataset of 2{,}438 tasks and yields only 500 trajectory samples for training. In contrast, SWE-Dev provides a large-scale repository-level training set with runnable environments and unit-test-based supervision, enabling scalable SFT, RL, and multi-agent training with direct execution feedback. In contrast, our proposed SWE-Dev provides a large-scale repository-level training set with runnable environments and test-based supervision. It supports SFT, RL, and MAS training with executable feedback, enabling realistic and scalable development of code generation systems.

\vskip -0.3in
\section{SWE-Dev}
\vskip -0.05in
SWE-Dev is the first dataset designed to train and evaluate autonomous coding systems on feature-driven software development tasks. Each instance requires the model to implement a new capability within an existing codebase, based on a natural language requirement and evaluated through real-world unit tests. 
This section describes the construction of the dataset (\S~\ref{sec:dataset-construction}), its core features (\S~\ref{sec:dataset-features}).

\begin{figure*}[t]
    \centering
    \includegraphics[width=\linewidth]{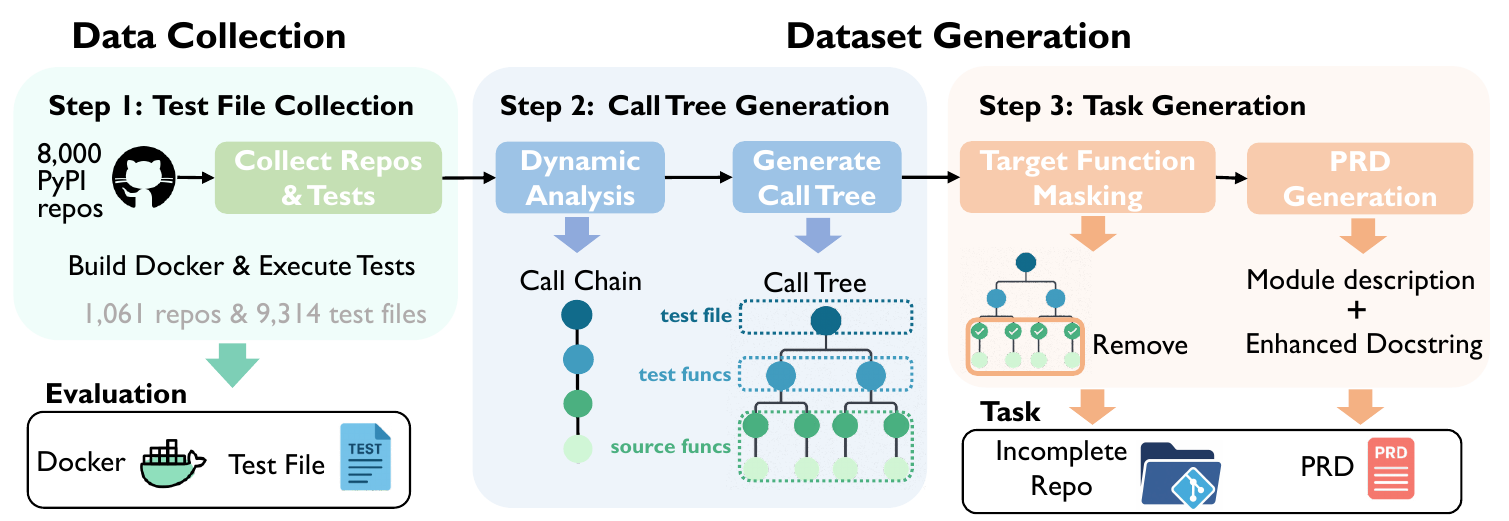}
    \caption{Overview of SWE-Dev dataset construction. \textbf{Step 1}: We collect real-world repositories with passing test files in Dockerized environments, \textbf{Step 2:} trace test executions to construct function-level call trees linking test cases to invoked source code, and \textbf{Step 3}: mask core functions while generating refined PRDs to create tasks. Each sample includes an incomplete repository, a natural language requirement, and executable test cases-enabling realistic, verifiable feature development.}
    \label{fig:enter-label}
\end{figure*}

\subsection{Dataset Construction}
\label{sec:dataset-construction}

Our dataset construction leverages a straightforward principle: test files in real-world repositories can serve both as a source of feature requirements and as a means of verifying correct implementation. 
In PyPI packages, developers create high-quality test files to ensure that specific modules or features function reliably across updates.
For example, in \texttt{numpy}, \texttt{test\_random.py} validates random array generation, aligning closely with the feature it tests. These test files provide executable, feature-specific validation, making them ideal for defining and evaluating development tasks.

Using these developer-authored tests as ground truth, we gain two advantages. First, they provide executable, functionality-level feedback for model evaluation. Second, by tracing the test cases back to their associated implementation code, we can identify and mask the relevant source code, forming the basis of an incomplete development task. These traced functions also guide the generation of precise task descriptions. Based on this process, we divide our construction into three stages: (1) collecting repositories, test files and building Docker environments, (2) generating test-to-code call trees via dynamic tracing, and (3) creating the final task by masking the relevant source code and producing the feature specification. 

\textbf{Step 1: Test File Collection}
To support realistic feature-level tasks and test-based evaluation, we begin by collecting real-world repositories that reflect common development scenarios. Specifically, we select 8,000 popular PyPI packages based on star counts. However, not all repositories are suitable: many lack usable tests or need sophisticated installation. Therefore, we applied a strict filtering process focused on test suite executability. Repositories were retained only if they met two criteria: (1) they include an identifiable test suite (e.g., using \texttt{pytest} or \texttt{unittest}), and (2) their test files could be executed successfully within the package Docker environment, with all tests passing. This ensures the resulting tasks are grounded in verifiable, runnable functionality. After filtering, we obtain 1,072 validated repositories (as of December 12, 2024) and 9,314 executable test files.

\textbf{Step 2: Call Tree Generation}
\label{sec: call tree}
To locate the specific methods involved in implementing a feature, we capture the runtime interactions between test cases and their corresponding source code through dynamic analysis. This process has two main parts:
(1) Dynamic analysis: We execute each test file using \texttt{pytest} inside a Docker environment and apply Python’s built-in \texttt{trace} module to record all triggered functions in source code. This results in multiple linear call chains that record the sequence of invoked source functions.
(2) Call tree ensemble: We aggregate the call chains into a hierarchical call tree, where the nodes of call tree represent functions, and edges capture dependency relationships. The call tree is rooted from test functions and followed by triggered source functions. The depth and width of the tree reflect the complexity of the feature logic, including nested structures and cross-file dependencies. These trees provide a precise mapping from test behavior to implementation code, enabling us to localize relevant functions and systematically control task difficulty later.

\textbf{Step 3: Task Generation}
Once we have localized the implementation logic using call trees, we transform it into a feature development task by (1) masking the existing implementations of feature components and (2) generating a natural language requirement for this feature. 
These components represent a typical development scenario where a new feature needs to be added to an existing system, but its implementation is yet to be realized. 
To achieve this, we perform the following: 
(1) Target function masking: We use structural properties of the call tree (e.g., depth and node count) to identify key function nodes that represent the core logic for the feature.  
The corresponding implementation code is removed from the repository, creating a space for the new feature to be integrated. 
(2) Project Requirement Document (PRD) generation: We construct the feature description in PRD by using GPT-4o to synthesize a high-level module description from the test file and augmenting the masked function’s docstring with implementation-level details. These two elements are combined into PRD, which serves as the task prompt. See example in Fig.~\ref {fig:example} and prompts in Appendix~\ref{app: prompt}.

\subsection{Dataset Features}
\label{sec:dataset-features}

\textbf{Controlled Complexity via Call Tree:} Leveraging call-tree analysis, SWE-Dev enables systematic adjustment of task difficulty by varying the masked dependency depth during task generation, supporting rigorous assessment under controlled complexity levels (see \S~\ref{abl:complexity}). We define test split \textit{Easy} tasks as those with call-tree depth 1--2, and \textit{Hard} tasks as those with depth $\geq 3$, which typically require implementing 5+ interdependent functions.

\textbf{Reliable Test-Based Evaluation:}
Assessment uses the original, developer-authored unit tests, validated for correct execution in a controlled environment. This execution-based pass/fail verification provides an objective, reproducible, and functionally accurate measure of code, directly reflecting real-world correctness criteria.

\textbf{Repository-Level Split for Clean Evaluation:}
SWE-Dev enforces a repository-level split between training and test sets to reduce contamination from shared repository context. The training split contains \textbf{1,033} repositories, while the test split contains \textbf{39} repositories, with no repository overlap between splits. More statistics are shown in \S~\ref{app:statics}.

\section{Experiment}
\vskip -0.05in
In this section, we empirically evaluate the effectiveness of various coding systems and training paradigms on SWE-Dev. We first compare the performance of single-LLM(\S~\ref{sec: test}) and MAS(\S~\ref{sec: multi test}) on the feature-driven software development tasks. Then, the effectiveness of different training approaches, including SFT (\S~\ref{sec: single sft}), RL (\S~\ref{sec: single rl}), and multi-agent training (\S~\ref{sec: multi train}) is discussed.

\textbf{Setup.} 
We employed the Pass@$k$ as an evaluation metric in SWE-Dev~\cite{humaneval}. 
For non-tool-use-agent inference code context, since SWE-Dev requires both the PRD and codebase as inputs. The codebases consist of many tokens (an average of \~20K lines, see Table~\ref{tab:statistics}), exceeding typical LLM context window. Thus, in all the experiments below, we provide only the relevant code context in the call tree rather than the entire codebase, since the call tree track exactly the code used for completion the task.

\subsection{Testing Results}
\label{sec: test}
This section presents the performance of 17 chatbot LLMs, 10 reasoning LLMs, 10 multi-agent systems, and 8 tool-augmented LLM agents (via OpenHands) on SWE-Dev, under the single-LLM, multi-agent, and agentic settings. Full details of the evaluated methods are provided in Appendix~\ref{app:inference}.

\subsubsection{Single LLM Inference}
\label{sec: single test}

\textbf{SWE-Dev presents substantial challenges for current LLMs, revealing a clear gap between existing coding capabilities and real-world software engineering demands.}

The left two part of Figure~\ref{fig:single llm performance} reports Pass@1 performance of chatbot and reasoning LLMs on SWE-Dev (see full results in Table~\ref{table: single inference}). We observe that: (1) LLMs perform better on the \textit{easy} split than the \textit{hard} split. 
(2) Performance generally scales with model size, especially for LLMs within the same family, aligning with our understanding of LLM capabilities. 
(3) Even the best-performing chatbot LLM: Claude-3.7-Sonnet~\cite{claude3.7}) achieves 19.74\% on the \textit{Hard} split and 53.09\% on the \textit{Easy} split. The reasoning LLM: Grok-3-beta achieves 53.63\% and 18.97\% on \textit{Easy} and \textit{Hard} split, respectively. This still falls short of achieving strong performance, indicating that current models are not yet fully capable of handling tasks that approximate the complexity of real-world development scenarios.

\subsubsection{Multi-Agent Inference}
\label{sec: multi test}

Table~\ref{tab: mas results} compares the performance, call times and total costs of various MAS against the single-agent baseline driven by GPT-4o-mini. 
Details of MAS are given in Appendix\ref{app: MAS}.

\textbf{MAS generally outperforms single-agent baselines on complex tasks.} 
While the single-agent approach achieves only 11.09\% Pass@1 on hard tasks, Self Refine~\cite{self_refine} and EvoMAC~\cite{evomac} improve performance to 20.03\% and 13.60\%, respectively. 
These results highlight the advantage of MAS in solving complex, reasoning-intensive problems.

\textbf{Simpler multi-agent strategies offer strong performance–efficiency trade-offs.}
Methods such as Self Refine strike an effective balance between performance and cost. On the easy subset, Self Refine achieves the highest Pass@1 of 40.02\% using only 5 calls. In contrast, more complex systems like ChatDev, despite making over 26 calls, fall behind in performance (35.13\%), indicating that additional agent complexity does not necessarily lead to better results.

\textbf{Human-designed, workflow-heavy MAS often introduce unnecessary overhead.} 
Systems with manually defined roles and interaction protocols, such as ChatDev and MapCoder, tend to be less effective. On hard tasks, ChatDev requires over 30 calls yet only achieves 11.7\%, while MapCoder performs even worse, with 5.87\% despite 23.41 calls. These results suggest that handcrafted workflows may introduce redundant operations without improving code generation quality.

Our results highlight MAS’s potential for complex tasks on SWE-Dev but reveal a gap between simple and complex MAS, indicating that scalable, efficient MAS remain a challenge. 
Future work could focus on balancing collaboration benefits with resource costs and mitigating error amplification from LLM hallucinations across agent interactions.

\subsubsection{LLM Agent Inference}
\label{sec: agent test}
 The right part in Figure~\ref{fig:single llm performance} reports the performance of tool-augmented \textbf{LLM agents} evaluated on SWE-Dev using the OpenHands~\cite{wang2024openhands} framework (see Table~\ref{tab:agent_results} in Appendix~\ref{app:llm_agent} for detailed number). Compared to direct single-LLM inference, agentic execution with tools and test feedback leads to substantial performance improvements.

\textbf{Agentic execution significantly improves feature development performance.}
Across all evaluated models, LLM agents consistently outperform their single-turn counterparts. Strong agents achieve over 50\% Pass@1 on the \textit{Hard} split (e.g., Claude4.1-Opus-thinking at 56.44\% and GPT-5 at 54.21\%), representing more than a 2$\times$ improvement over the best single-LLM results. This demonstrates the importance of iterative execution, error diagnosis, and repair when tackling repository-level feature development.

\textbf{Reasoning-enabled agents are strong under execution feedback.}
With OpenHands, reasoning-enabled (Thinking) agents achieve high performance on the \textit{Hard} split (e.g. Kimi-K2-Thinking is comparable to Kimi-K2-Instruct (50.86\% vs.\ 51.05\% on \textit{Hard} split), suggesting that the benefits of explicit reasoning are context-dependent and become most meaningful when coupled with execution signals.

\textbf{Despite strong gains, feature development remains unsolved.}
Even with agentic scaffolding, a large fraction of tasks on the \textit{Hard} split remain unsolved, highlighting the inherent difficulty of end-to-end feature development and leaving substantial headroom for future advances in agent design, planning, and long-horizon reasoning.

\subsection{Training Results}
\label{sec: train}
In this section, we evaluate SWE-Dev's support for different training methods, including SFT, RL. 
Additionally, we present preliminary results from our exploration of multi-agent training, offering an initial assessment of MAS-based learning. For detailed training setups, refer to the Appendix~\ref{app: train}.

\subsubsection{Single LLM SFT}
\label{sec: single sft}
We conducted experiments on Qwen2.5-Instruct models of various sizes (0.5B, 1.5B, 3B, and 7B) to assess the impact of SFT on performance in SWE-Dev. Experimental setting is in Appendix~\ref{app: train sft}.

\textbf{Training significantly improves performance across model sizes.}
SFT leads to substantial performance improvements across all model sizes, especially for harder tasks. As shown in Table~\ref{tab: single sft}, the 7B model achieves a Pass@1 of 36.90\% on the easy task set after fine-tuning, up from 25.74\% in the zero-shot setting. On the hard task set, the Pass@1 increases from 6.68\% to 18.89\%, demonstrating the clear benefits of training in enhancing model performance.

\begin{figure}[t]
    \centering
    \includegraphics[width=0.6\linewidth]{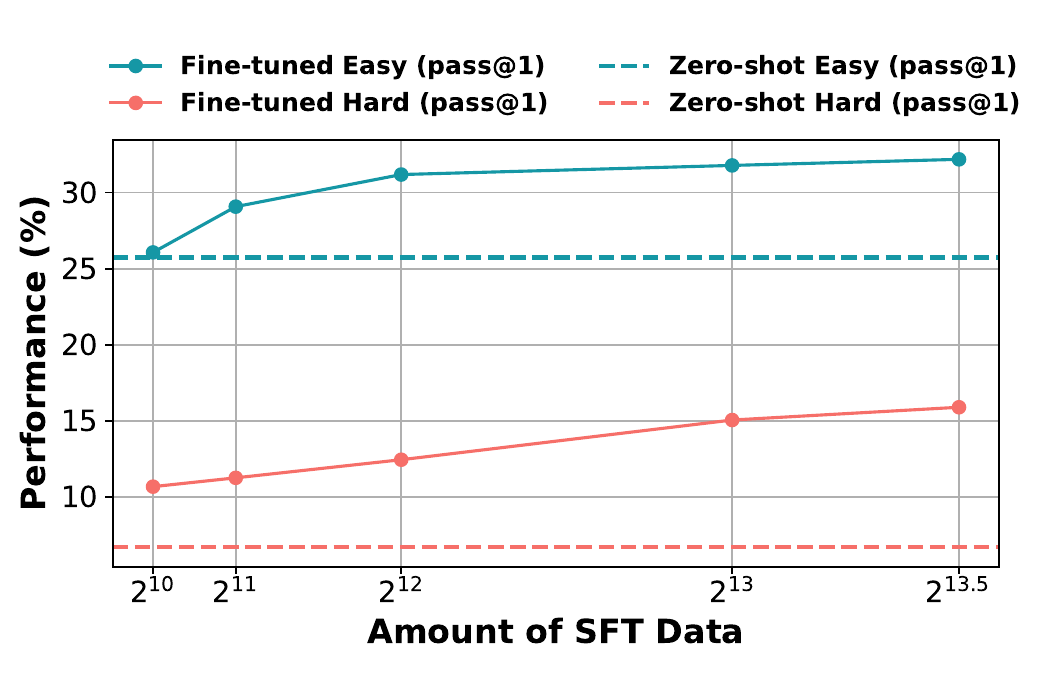}
    \caption{Training data scaling of SFT Qwen2.5-7B-instruct on SWE-Dev. As data size increases, performance improves steadily under SFT.}
    \label{fig: sft scaling law}
\end{figure}

\textbf{SWE-Dev effectively supports the scaling law of training.}
Figure~\ref{fig: sft scaling law} illustrates the scaling law of training using Qwen2.5-7b-instruct. In this experiment, we measured model performance across varying amounts of fine-tuning data, specifically tracking changes in Pass@1 for both easy and hard task. As shown in the figure, performance improves steadily as the amount of fine-tuning data increases, with larger improvements observed for harder tasks.

In summary, our results underscore the importance of fine-tuning in improving performance on SWE-Dev. 
The scaling law observed here further supports the idea that SWE-Dev is a valuable dataset for studying the effects of model size and training data on task performance.

\subsubsection{Single LLM RL}
\label{sec: single rl}

\begin{table}[t]
\centering

\caption{Performance comparison of Qwen2.5-7B-Instruct as
base model, SFT-Tuned and RL-Tuned models on SWE-Dev.}
\scalebox{0.9}{
\begin{tabular}{c|cc|cc}
\toprule
& \multicolumn{2}{c|}{\textbf{Pass@1}}                                 & \multicolumn{2}{c}{\textbf{Pass@3}}                          \\ \cmidrule(l){2-5} 
\multirow{-2}{*}{\textbf{Method}}     & \multicolumn{1}{c}{Easy}      & \multicolumn{1}{c|}{Hard}  & \multicolumn{1}{c}{Easy} & \multicolumn{1}{c}{Hard} \\ \midrule
\multicolumn{1}{c|}{\textbf{vanilla}} & 25.74                         & \multicolumn{1}{c|}{6.68}  & 33.35                    & 7.73                     \\
\multicolumn{1}{c|}{\textbf{SFT}} & 27.09                         & \multicolumn{1}{c|}{9.77}  & \textbf{34.49}                    & 13.63                     \\
\multicolumn{1}{c|}{\textbf{PPO (online RL)}}     & \textbf{28.30}                         & \multicolumn{1}{c|}{\textbf{12.25}} & 32.69                    & 14.33                    \\
\multicolumn{1}{c|}{\textbf{DPO (offline RL)}}     & 25.89                         & 10.36                      & 31.32                    & \textbf{14.66}                    \\ \bottomrule
\end{tabular}}
\label{tab: rl results}
\end{table}

SWE-Dev provides precise test cases enabling accurate rewards for coding tasks, supporting both online and offline RL. In this section, we explore the impact of RL on the Qwen2.5-7B-instruct using SWE-Dev. 
Considering the computational cost of RL, we limit our experiments in this section to 2k training samples.
For full training setup, refer to the Appendix~\ref{app: train rl}.

\begin{table}[t]
    \centering
    \caption{Comparison of zero-shot and SFT performance (Pass@1) on SWE-Dev using Qwen2.5 models.
    Results are reported on both Easy and Hard test splits across model sizes from 0.5B to 7B. The $\Delta$ columns indicate relative improvement after SFT. Fine-tuning yields consistent gains.}
    \scalebox{0.9}{
    \begin{tabular}{l|cc|cc|cc}
    \toprule
     & \multicolumn{2}{c|}{\textbf{Zero-shot}} & \multicolumn{4}{c}{\textbf{SFT}} \\
     \cmidrule{2-7}
    \multirow{-3}{*}{\textbf{}} & Easy & Hard & Easy & $\Delta$(\%) & Hard & $\Delta$(\%) \\ \midrule
    \textbf{0.5B} & 6.39 & 1.00 & 12.12 & +90 & 4.37 & +337 \\
    \textbf{1.5B} & 8.05 & 1.23 & 18.20 & +126 & 7.64 & +521 \\
    \textbf{3B} & 15.93 & 5.27 & 27.53 & +73 & 12.46 & +136 \\
    \textbf{7B} & 25.74 & 6.68 & 36.90 & +43 & 18.89 & +183 \\ \bottomrule
    \end{tabular}}
    \vskip -0.1in
    \label{tab: single sft}
\end{table}

\textbf{Both online and offline RL improve performance, especially on hard tasks.} 
Table~\ref{tab: rl results} shows that both PPO~\cite{ppo} and DPO~\cite{dpo} significantly improve Pass@1 performance, especially on the \textit{Hard} split.
Furthermore, PPO outperforms SFT~(2k, aligned with RL) on the same training data. 
These findings highlight the advantages of RL training.

\textbf{RL boosts one-shot success but not multi-sample gains.}
While RL fine-tuning yields improvements in Pass@1, it has minimal impact on Pass@3. Specifically, PPO achieves a Pass@1 of 28.30\% on easy tasks, a noticeable increase from the base model’s 25.74\%, but the Pass@3 remains lower than the SFT-training, even the original model's performance.
These results suggest that RL can be beneficial in refining Pass@1, particularly for more complex tasks, by increasing the model's efficiency in generating correct answers in fewer attempts. 
However, this efficiency comes at the cost of reduced exploration. 
This aligns with findings from prior work~\cite{yue2025doesreinforcementlearningreally}. 
Therefore, while RL improves performance, significant headroom remains, and more advanced methods or further training are needed to achieve improvements across tasks.

\begin{table}[t]
        \centering
        \caption{Comparison of multi-agent role-wise training, base MAS and single LLM's performance on Qwen2.5-7B-Instruct. $\Delta$ indicates the relative improvement over the base MAS system. Partial fine-tuning of either agent also leads to consistent gains, demonstrating the effectiveness of role-specific supervision enabled by SWE-Dev. \textit{FT} and \textit{base} refers to finetuning and non-finetuning.}
        \scalebox{0.9}{
        \begin{tabular}{l|cc|cc|cc}
            \toprule
            & \textbf{Org} & \textbf{Coder} & \textbf{Easy} & $\Delta$(\%) & \textbf{Hard} & $\Delta$(\%) \\ \midrule
            \textbf{Single} & - & - & 25.74 & - & 6.68 & - \\ \midrule
\multirow{4}{*}{\textbf{MAS}} & base & base & 26.64 & - & 7.39 & - \\
 & FT & base & 30.04 & +12.76 & 12.36 & +67.25 \\
 & base & FT & 31.42 & +17.94 & 11.49 & +55.48 \\
 & FT & FT & \textbf{31.65} & +18.80 & \textbf{12.70} & +71.85 \\ \bottomrule
        \end{tabular}}
        \label{tab: mas training}
\end{table}

\begin{figure*}[t]
    \begin{minipage}[t]{0.33\textwidth}
        \centering
        \includegraphics[width=\linewidth]{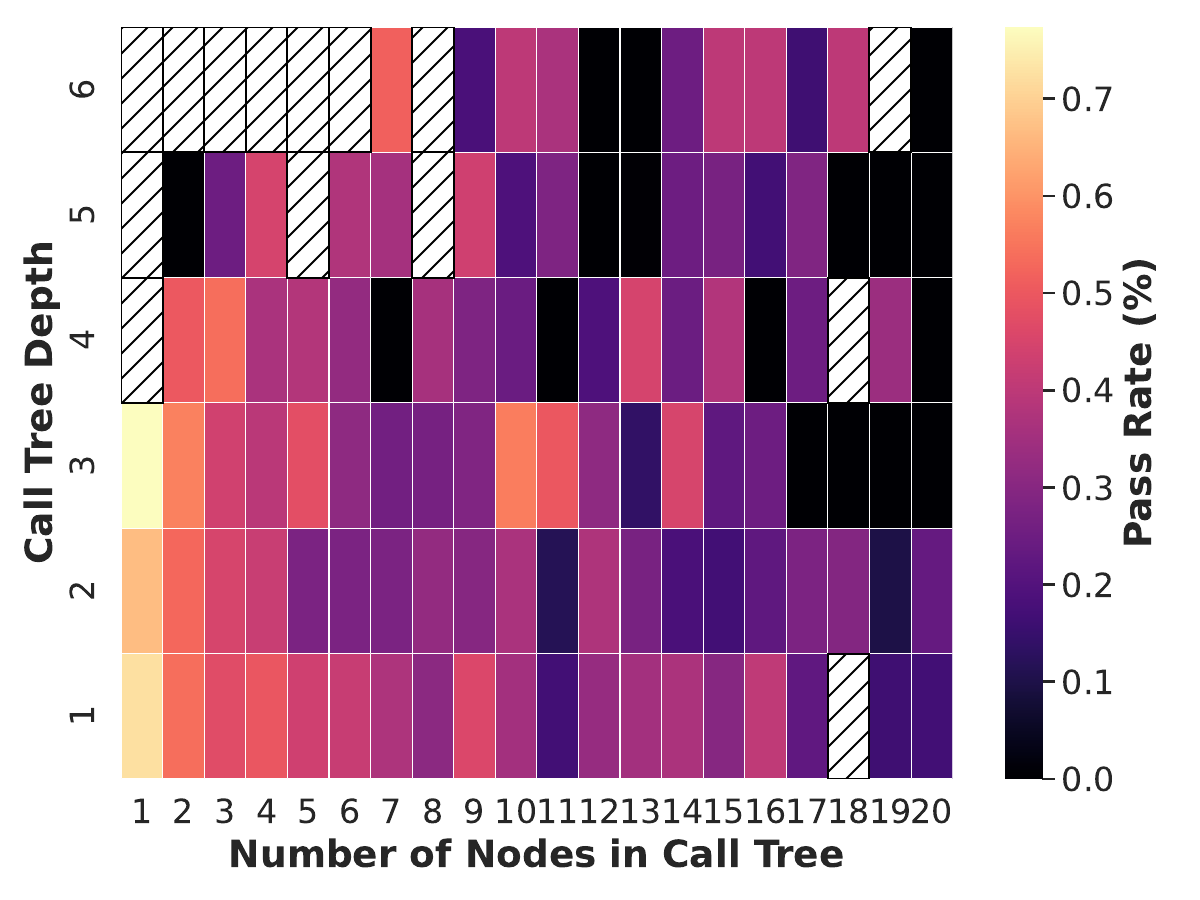}
        \subcaption{Complexity analysis}
        \label{fig:nodes_vs_depth}
    \end{minipage}
    \begin{minipage}[t]{0.32\textwidth}
        \centering
        \includegraphics[width=\linewidth]{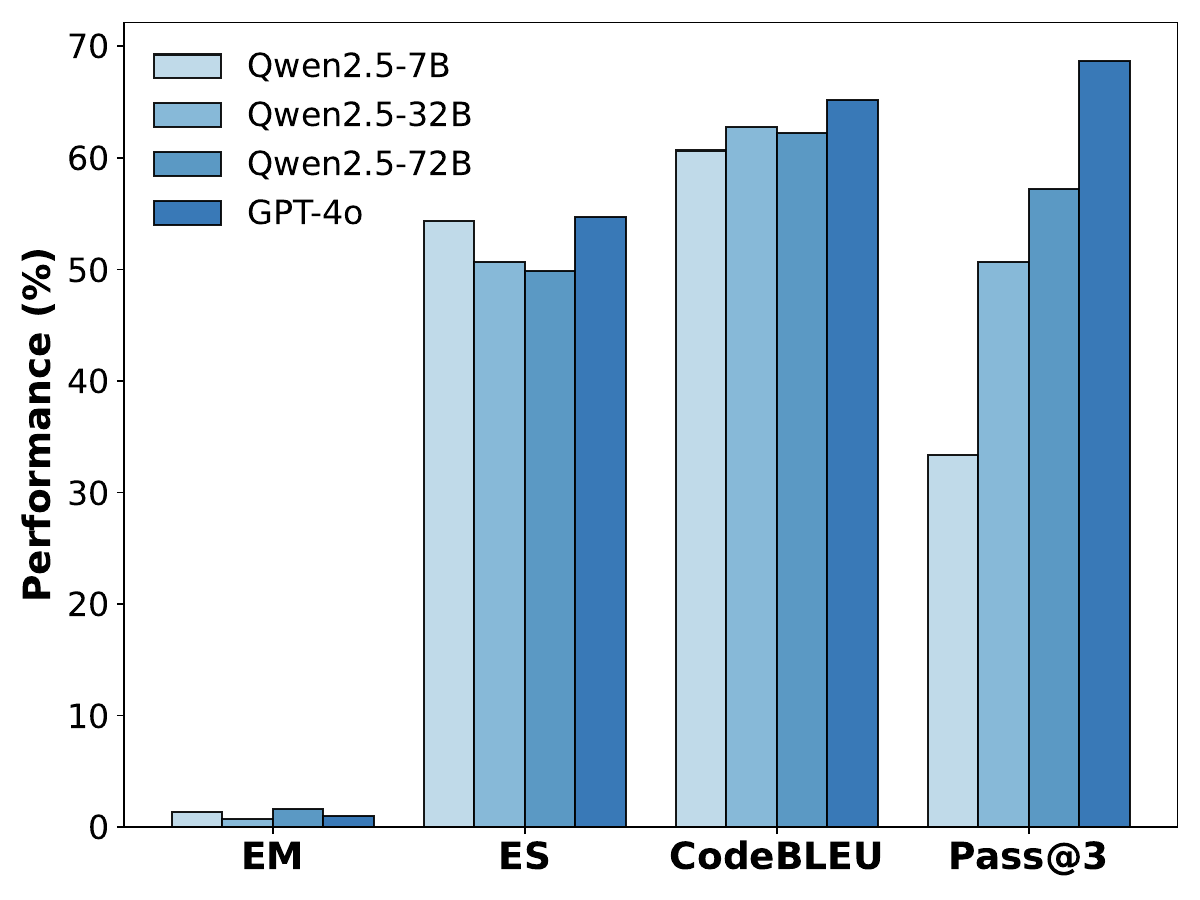}

        \subcaption{Metric Comparison}
        \label{fig:testcase_eval}
    \end{minipage}
    \begin{minipage}[t]{0.32\textwidth}
    \centering
    \includegraphics[width=\linewidth]{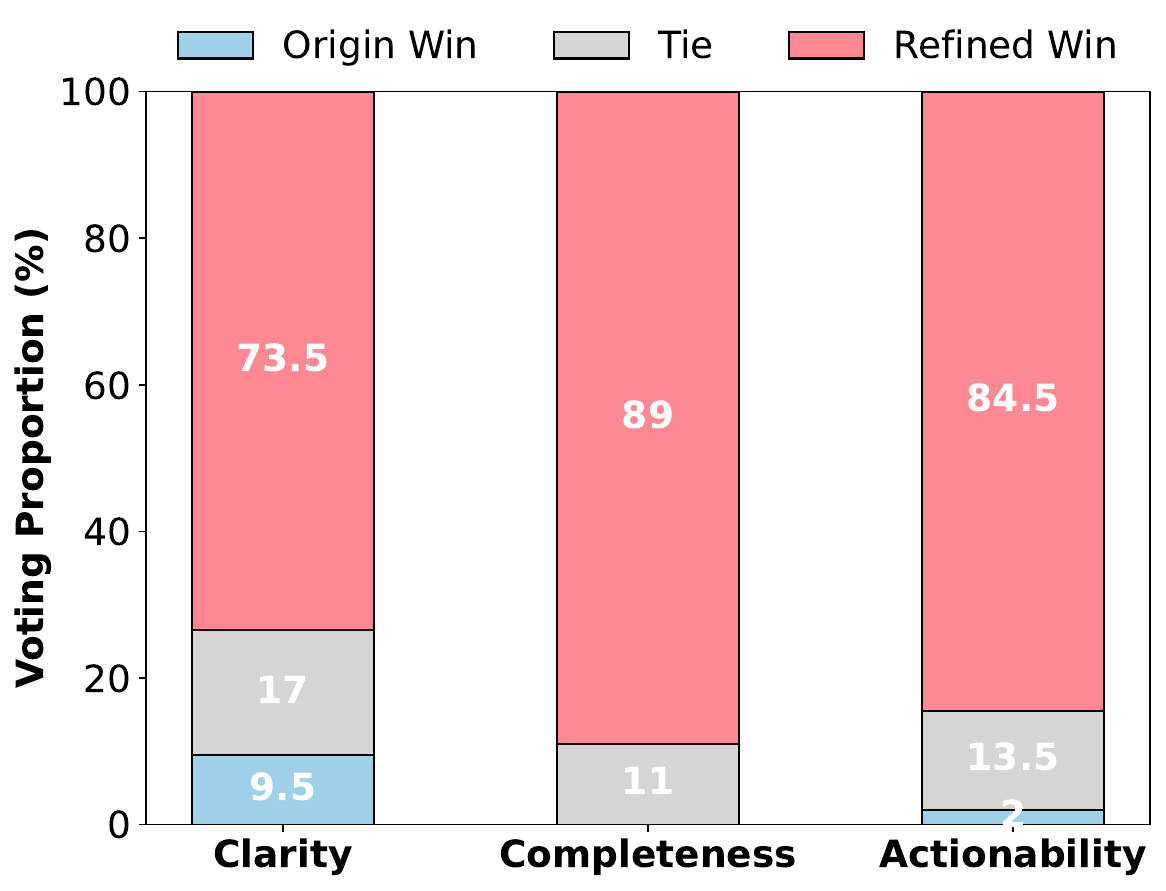}
    \subcaption{PRD Quality Analysis}
    \label{fig:prd_comparison_}
    \end{minipage}

    \caption{Analysis of SWE-Dev Benchmark Characteristics. (a) Compares GPT-4o’s performance across tasks grouped by call tree depth and node count, showing that greater structural complexity correlates with lower accuracy. (b) Compares several evaluation metrics; Pass@3 shows the clearest differentiation across model scales. (c) Compares human ratings of original vs. refined PRD on 100 samples of 3 dimensions, revealing SWE-Dev's high PRD quality.}
\end{figure*}

\begin{figure}[t]
    \centering
    \includegraphics[width=0.5\linewidth]{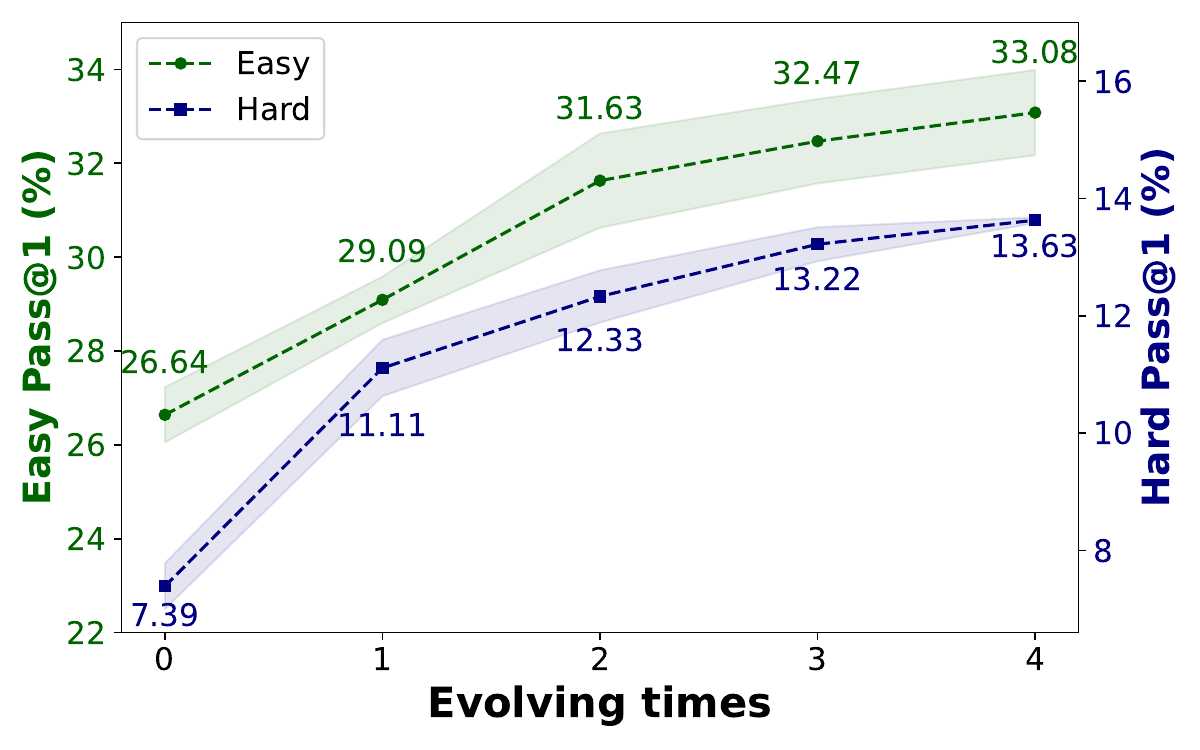}
    \vskip -0.1in
    \caption{EvoMAC performance trajectory under ground truth test case supervision on SWE-Dev with Qwen2.5-7B-Instruct. EvoMAC iteratively improves across reasoning rounds, guided by executable test feedback.}
    \vskip -0.2in
    \label{fig:evolve}
\end{figure}

\subsubsection{Multi-Agent Training}
\label{sec: multi train}
MAS has shown promising results on SWE-Dev, and we further investigate the training process of MAS on this dataset. As depicted in Fig.~\ref{fig:evolve}, the ground truth test case supervision in SWE-Dev enables EvoMAC~\cite{evomac} to improve its performance across multiple rounds of reasoning. This iterative refinement process motivates us to explore EvoMAC as the MAS for training in SWE-Dev. We apply rejection sampling to enhance agent performance via role-wise training.

\textbf{Trajectory Collection.}
We use Qwen2.5-7B-Instruct to collect training data for the MAS, following these steps: 
(1) \textbf{EvoMAC iterative reasoning}: EvoMAC performs multiple reasoning rounds, benefiting from ground truth test case supervision to progressively improve its performance. 
(2) \textbf{Rejection sampling}: At each iteration, we apply rejection sampling based on training sample testcases to select high-quality trajectories that show improvement over the previous round, ensuring the retention of beneficial data.
(3) \textbf{Role-wise training}: The selected trajectories are used to role-wise train two agents (organizer and coder) in EvoMAC, allowing each agent to specialize in its task for better overall performance.

\textbf{Training Effectiveness.} 
Table~\ref{tab: mas training} presents the performance of different training configurations in terms of Pass@1. 
We see that: i) Fine-tuning both the organizer and coder agents results in the highest performance, with Pass@1 of 31.65\% on easy tasks and 12.70\% on hard tasks, outperforming all other configurations; ii) When only one agent is fine-tuned, we also see improvements over the baseline. 
These findings highlight the effectiveness of role-wise training for MAS training.

\section{Dataset Analysis}
\vskip -0.05in

We analyze SWE-Dev’s task complexity, evaluation setup, and PRD quality to demonstrate its uniqueness and reliability.
\noindent \textbf{Analysis of Task Difficulty and Call Tree Characteristics.} 
\label{abl:complexity} We analyze how task difficulty in SWE-Dev correlates with call tree complexity. As introduced in \S~\ref{sec: call tree}, a call tree reflects the dynamic function invocation structure for this task, where nodes represent functions and edges denote their call relationships. We use two metrics: depth, indicating the maximum call nesting, and node count, representing the total number of distinct functions involved in the task. Fig.~\ref{fig:nodes_vs_depth} shows that GPT-4o's performance declines as depth and node count increase, revealing a strong correlation between structural complexity and task difficulty. This suggests that deeper and broader call structures introduce more functional requirements and interdependencies, making tasks more challenging.

\noindent \textbf{Evaluation Method Precision.} 
\label{sec: eval abl}
SWE-Dev uses execution-based evaluation with test cases, enabling precise performance signals. We compare metrics: Exact Match (EM)~\cite{m2rc-eval}, Exact Sequence (ES)~\cite{m2rc-eval}, CodeBLEU~\cite{codebleu}, and Pass@3, using Qwen2.5 models and GPT-4o. As Fig~\ref{fig:testcase_eval} shows, Pass@3 best reflects capability scaling, separating models by size and quality. In contrast, EM, ES, and CodeBLEU show minimal variance, failing to distinguish models. This demonstrates that SWE-Dev’s test-case-based evaluation provides a more robust and realistic signal of model performance, better reflecting the functional correctness required in real-world software development.

\noindent \textbf{PRD Quality.} 
SWE-Dev includes a PRD for each task to simulate realistic developer-facing requirements, which are primarily derived from the original docstrings found within the repository source code. 
While many functions in open-source code include docstrings, we found that these are often incomplete—lacking clear descriptions of behavior, parameters, or edge cases. To improve instruction clarity without fabricating content, we lightly refine existing docstrings using GPT-4o, grounded in the related file and surrounding context. To evaluate instruction quality, we conducted a human assessment on 100 sampled tasks. Two experienced engineers rated the original and refined PRDs along Actionability, Completeness, and Clarity (Appendix~\ref{app: prd_quality analysis} includes human instruction). As shown in Fig.~\ref{fig:prd_comparison_}, refined PRDs consistently scored higher across all dimensions. This supports SWE-Dev’s goal of providing realistic, well-scoped requirements for reliable model evaluation.

\vskip -0.1in
\section{Conclusion}
\vskip -0.05in
In this work, we introduced SWE-Dev, the first dataset for evaluating and training autonomous coding systems on end-to-end feature-driven development task. SWE-Dev consists of 14,000 training and 500 test instances, each uniquely equipped with runnable environments and developer-authored executable unit tests, which provides essential execution-based feedback for advanced training paradigms like SFT, RL, and multi-agent learning. Our experiments show feature-driven software development is profoundly challenging for current autonomous coding systems. We also validate that training on SWE-Dev can yield encouraging performance gains. These findings validate SWE-Dev as a critical platform for identifying current limitations and fostering breakthroughs in AI-driven software development. We hope the release of SWE-Dev spurs innovation in long-context reasoning, agent orchestration, and execution-aware training towards genuinely autonomous software engineering.

\newpage


\nocite{langley00}

\bibliography{main}

\newpage
\appendix
\onecolumn


\section{LLM Usage Disclosure}
In our work, we utilized GPT-4o exclusively to enhance readability and improve language fluency during the writing process. We take full responsibility for the content of this publication, including any content generated by LLM.

\section{Other Related Work}
\textbf{RL for Software Engineering}
Earlier RL and RLHF work on code has mostly targeted short, function-level problems (e.g., HumanEval-style benchmarks), where the model produces a single solution in one shot with an immediate test-based reward. More recently, several works have applied RL to more complex SWE settings~\cite{golubev2025training}. SWE-RL~\cite{swe-rl} uses policy-gradient RL in an agentless scaffold on SWE-Bench, but still frames the task as single-turn patch generation, which sidesteps the challenges of stateful, multi-step interaction and long-horizon credit assignment. In contrast, DeepSWE~\cite{deepswe} and SkyRL-v0~\cite{skyrl} scale RL to interactive, multi-turn SWE agents operating over full repositories, tools, and long contexts. Our work is complementary: rather than proposing a new agent or RL algorithm, SWE-Dev provides a large, PRD-driven, feature-level benchmark with executable rewards and examines how standard SFT and PPO behave on multi-file feature implementation. This makes SWE-Dev a natural substrate for future RL-based SWE agents in the spirit of these recent works.

\section{Dataset}
\label{app:statics}

\label{app: data available}

\subsection{Dataset Statistics}

\label{sec:dataset-statistics}

\begin{figure}[h]
\centering
\begin{minipage}{0.44\textwidth}
\centering
\includegraphics[width=\linewidth]{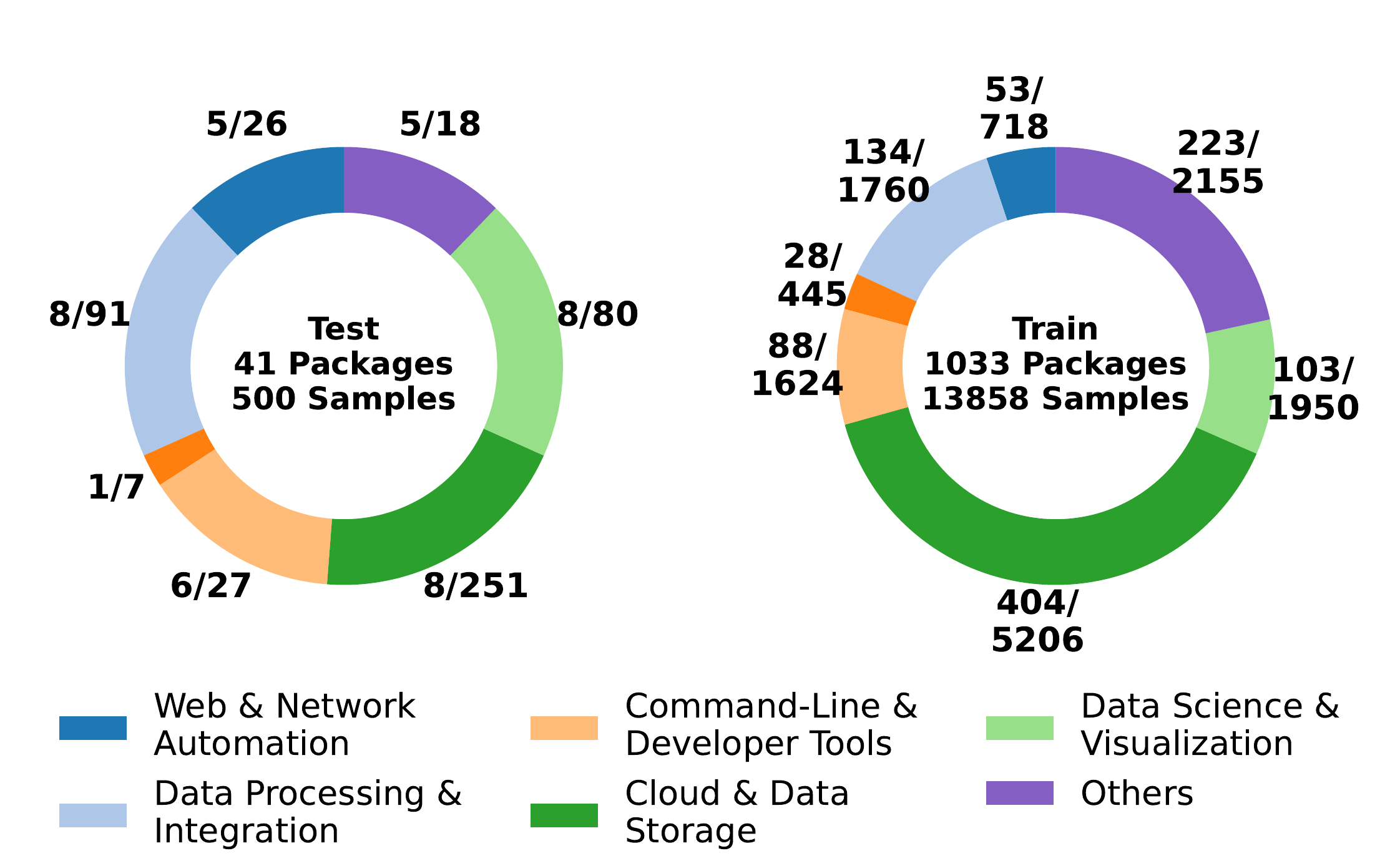}
\caption{Distribution of SWE-Dev training and test samples across 6 major PyPI application domains.}
\label{fig:dist_compare}
\end{minipage}
\begin{minipage}{0.55\textwidth}
\captionof{table}{Basic statistics of SWE-Dev, including task specification length, repository scale, ground truth implementation size, and evaluation test coverage for both train and test splits.}
\label{tab:statistics}
\centering
\scalebox{0.8}{
\begin{tabular}{l|l|cc|c}
\toprule
\multirow{2}{*}{\textbf{Category}} & \multirow{2}{*}{\textbf{Metric}} & \multicolumn{2}{c|}{\textbf{Test}} & \multirow{2}{*}{\textbf{Train}} \\ 
 &  & Easy & Hard &  \\ \midrule
\textbf{Size} & \# Samples & 250 & 250 & 14000 \\
 & \# Total repos & 39 & 39 & 1033 \\ \midrule
\textbf{Task} & \# Avg. tokens & 1499 & 2148 & 1833 \\ \midrule
\textbf{Codebase} & \# Avg. Files & \multicolumn{2}{c|}{71.31} & 64.40 \\
 & \# Avg. LOC & \multicolumn{2}{c|}{21308} & 20206 \\ \midrule
\textbf{GT Code} & \# Avg. LOC & 109.1 & 172.4 & 199.92 \\
 & \# Avg. funcs & 4.75 & 6.972 & 6.03 \\ \midrule
\textbf{Tests} & \# Avg. test lines & 134.8 & 123.9 & 90.9 \\
 & \# Avg. testcases & 6.62 & 4.29 & 5.92 \\ \bottomrule
\end{tabular}}
\end{minipage}

\end{figure} 

Table~\ref{tab:statistics} summarizes the key statistics of SWE-Dev, which consists of 14,000 training and 500 test samples. 
The test set is manually curated and split into two difficulty levels: \textit{easy} and \textit{hard} (250 instances each). Each dataset instance comprises four components: (1) the task, specified by a PRD, with its token count reflecting instruction length; (2) the codebase, referring to the non-test portion of the repository, where we report the number of files and lines of code (LOC); (3) the ground truth (GT) code to be implemented, measured by its LOC and number of target functions; and (4) the test suite, evaluated via the number of test cases and total test LOC per sample. Figure~\ref{fig:dist_compare} shows the distribution of training and test samples across six major PyPI application domains, demonstrating the diversity of software categories represented in the dataset. More statistics are in Appendix \ref{app:statics}.

\subsection{Dataset Information}
\begin{figure}[h]
    \centering
    \includegraphics[width=\linewidth]{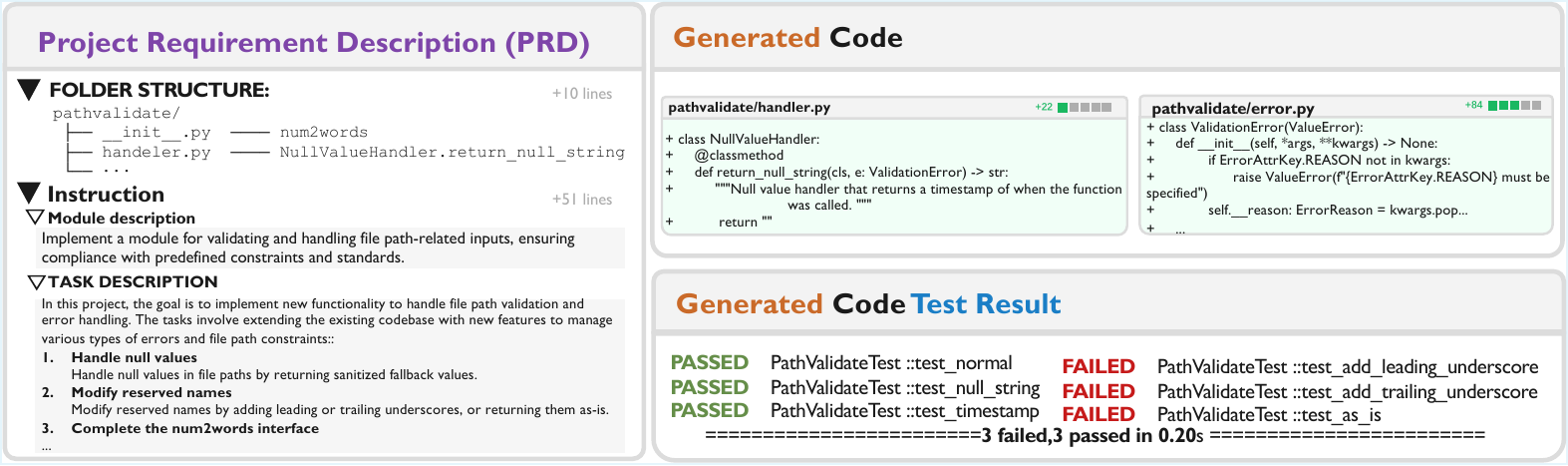}
    \caption{Example of a SWE-Dev Sample. Each sample includes a Project Requirement Description (PRD) with folder structure, module-level task description, and masked docstrings; the corresponding ground truth implementation. Generated code is evaluated by the test function execution results. This structure supports realistic, testable feature development in a repository context.}
    \label{fig:example}
\end{figure}

Figure~\ref{fig:example} illustrates a typical task instance in the SWE-Dev, detailing the entire development workflow.
The process begins with the Project Requirement Description (PRD), which provides instructions and specifies features to be implemented. Methods to be evaluated then generate code to complete the features mentioned in the PRD, which is subsequently verified against the test suite to produce pass/fail results to calculate pass rate. 
Additionally, the ground truth implementation for each PRD is included for reference. 
The tasks in SWE-Dev simulate real-world software development cycles within a repository context. For detailed information about each data field included in SWE-Dev tasks, please refer to Table~\ref{tab:swd_bench_fields_description_en}.

\begin{table*}[h]
\centering
\caption{Description of each field of an SWE-Dev task instance object.}
\begin{tabular}{l|p{0.7\textwidth}} 
\toprule
Field & Description \\
\midrule
\texttt{PRD} & A natural language document describing the project requirements, including the specific features to be implemented. \\
\texttt{file\_code}  & Incomplete code contents of the core files involved in the task. \\
\texttt{test\_code}  & Content of the test code used to verify the task's functionality. \\
\texttt{dir\_path}  &  Root directory path of the project corresponding to this task instance. \\
\texttt{package\_name} & Name of the software package or module to which this task instance belongs. \\
\texttt{sample\_name} & Unique identifier or name for this task instance or sample within the benchmark. \\
\texttt{src\_dir} & Relative directory path where the source code files for the project or task are located. \\
\texttt{test\_dir} & Relative directory path where the test code files for the project or task are located. \\
\texttt{test\_file} & Relative path of the unit test file used for executing tests. \\
\texttt{GT\_file\_code} & Ground Truth source code for the file to complete. \\
\texttt{GT\_src\_dict} & Ground Truth source dictionary, mapping file names/paths to their expected correct code content. \\
\texttt{dependency\_dict} & Dictionary describing the dependencies required by the current task (e.g., internal modules) and their relationships. \\
\texttt{call\_tree} & Function call tree or call graph of the code, representing the relationships between function calls. \\
\bottomrule
\end{tabular}
\label{tab:swd_bench_fields_description_en} 
\end{table*}

\subsection{Dataset Distribution}
We present the distribution statistics of the training and test sets in SWE-Dev. Each sample includes a Project Requirement Document (PRD), which describes the feature to be implemented. The average PRD length is 1,845.4 tokens. On average, each sample includes at least 5 unit tests for functional evaluation, spans 3 source files, and requires the implementation of approximately 6 functions.

\begin{figure}[h]
    \centering
    \begin{minipage}[t]{0.48\textwidth}
        \centering
        \includegraphics[width=\linewidth]{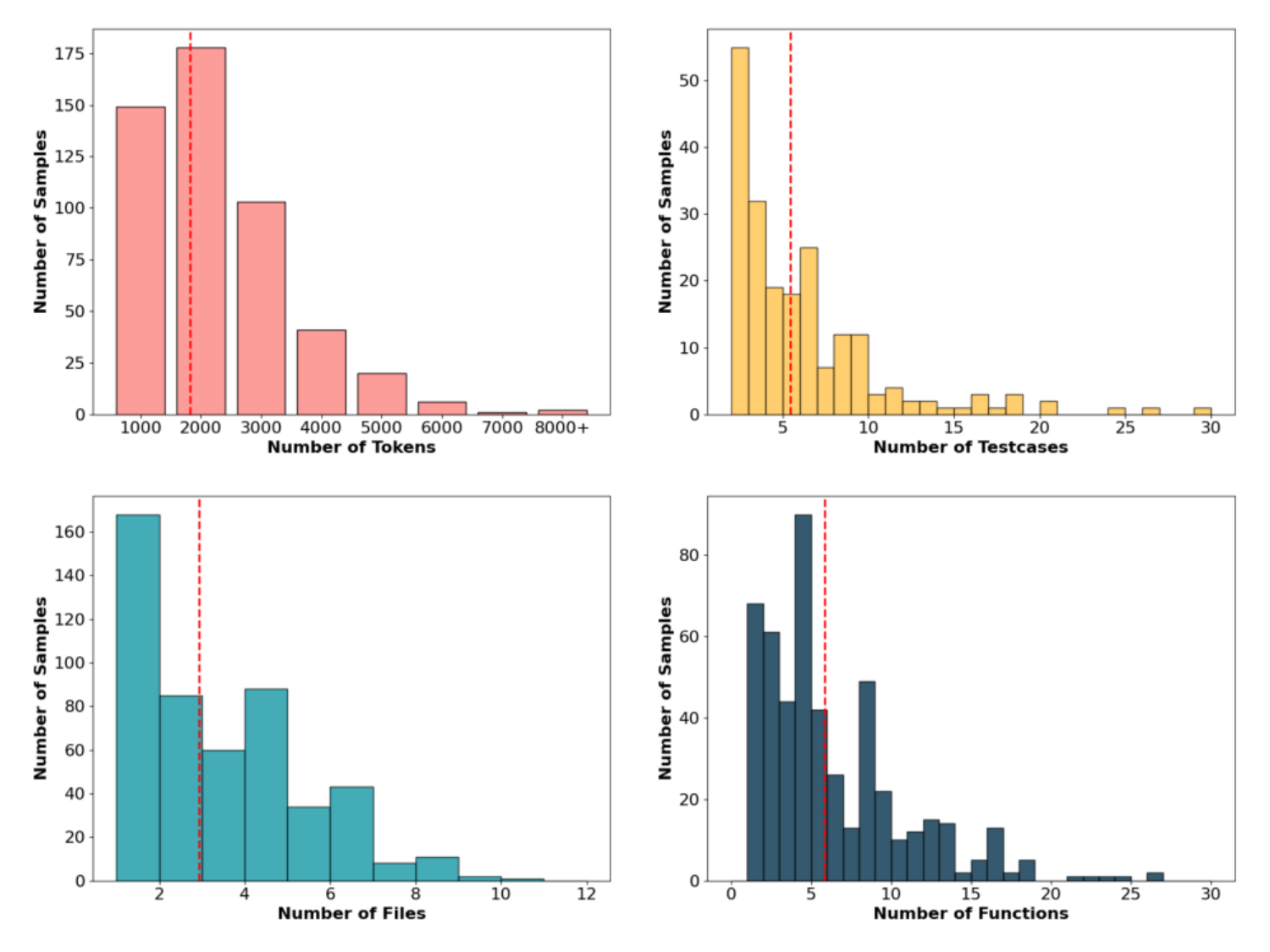}
        \subcaption{Test Set Distribution}
        \label{fig:enter-label}
    \end{minipage}
    \begin{minipage}[t]{0.48\textwidth}
        \centering
        \includegraphics[width=\linewidth]{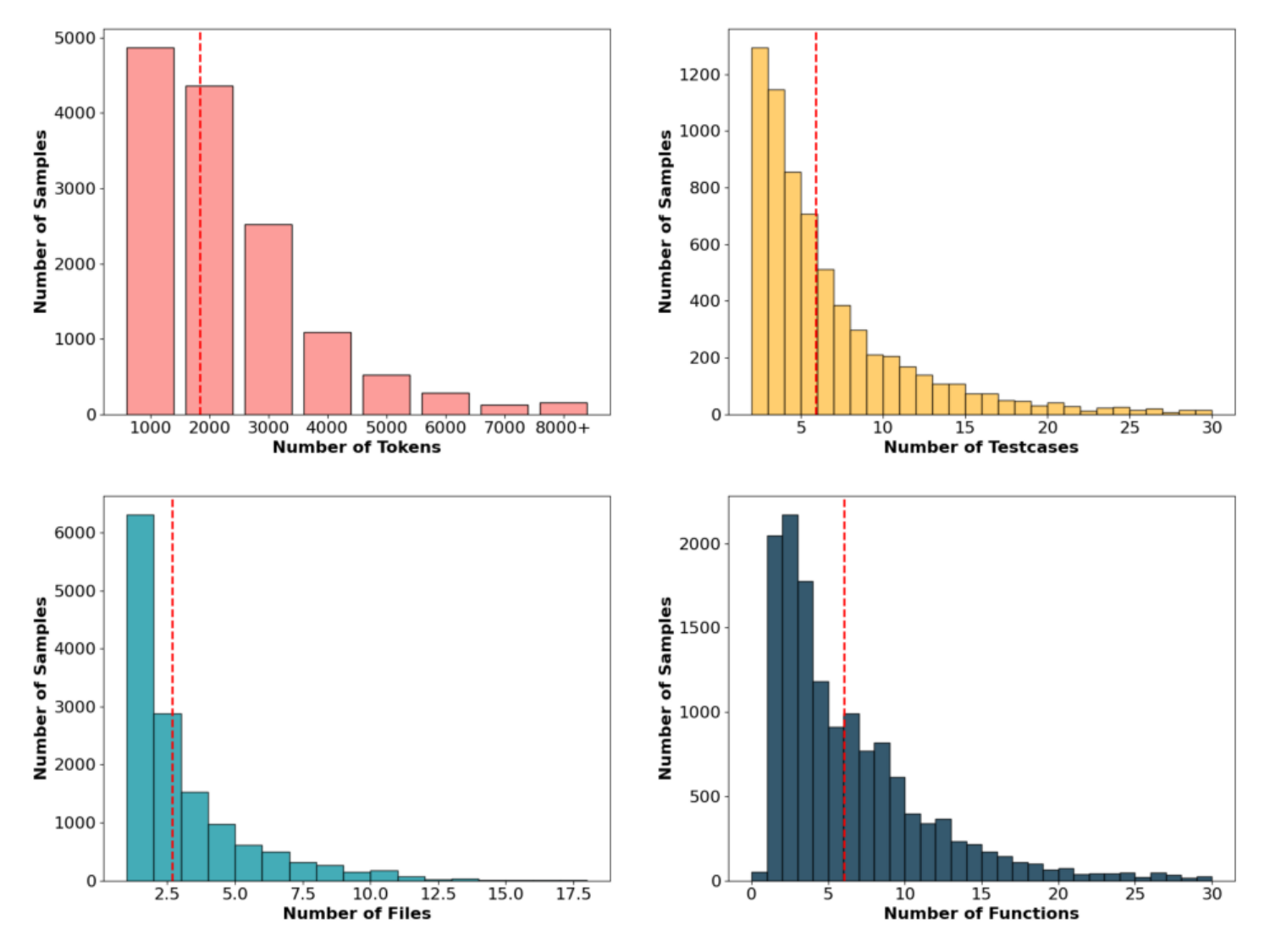}
        \subcaption{Train Set Distribution}
        \label{fig:enter-label}
    \end{minipage}
    \caption{Dataset distribution of PRD tokens, number of testcases, number of files to complete, and number of functions per sample.}
\end{figure}



\subsection{Dataset Diversity}

\begin{figure}[h]
    \centering
    \includegraphics[width=0.8\linewidth]{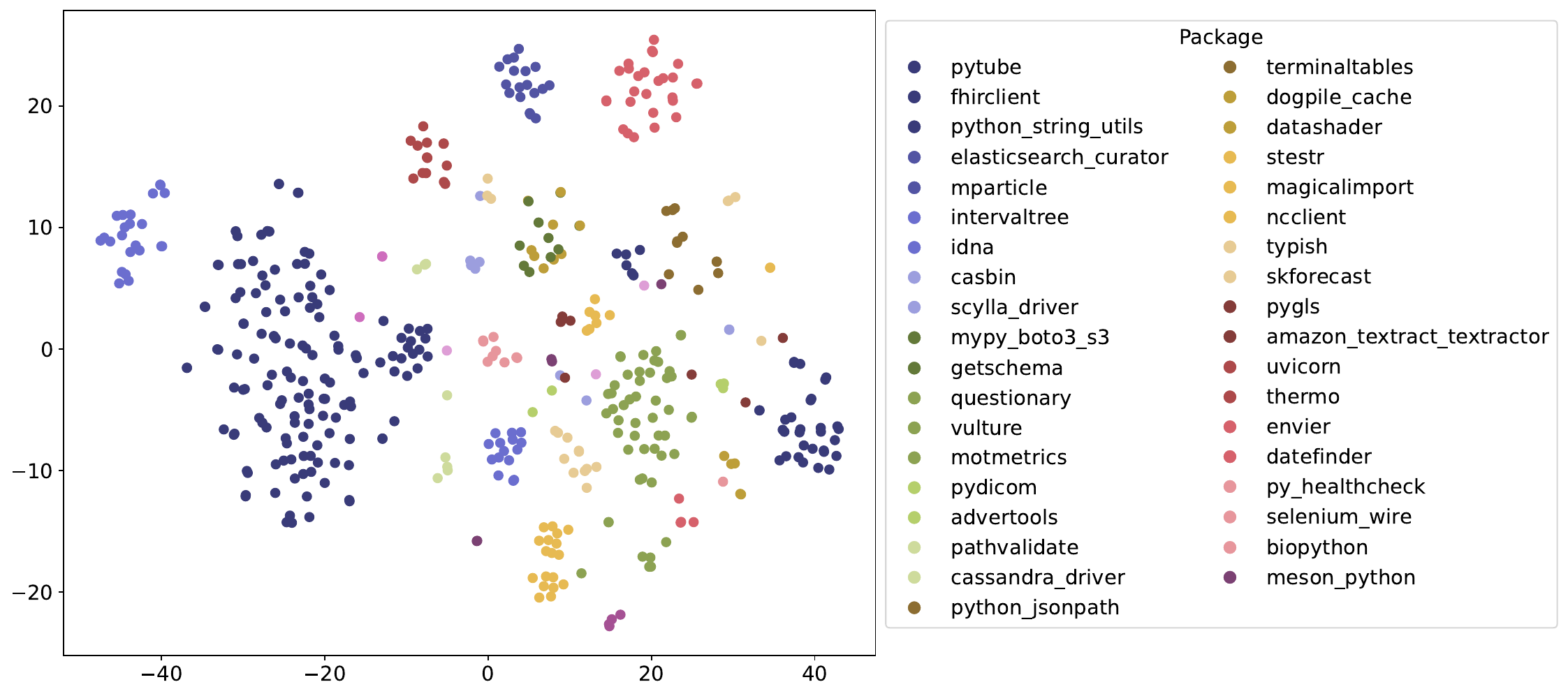}
    \caption{t-SNE visualization of PRD in test set}
    \label{fig:tsne test}
\end{figure}

\begin{figure}[h]
    \centering
    \includegraphics[width=0.8\linewidth]{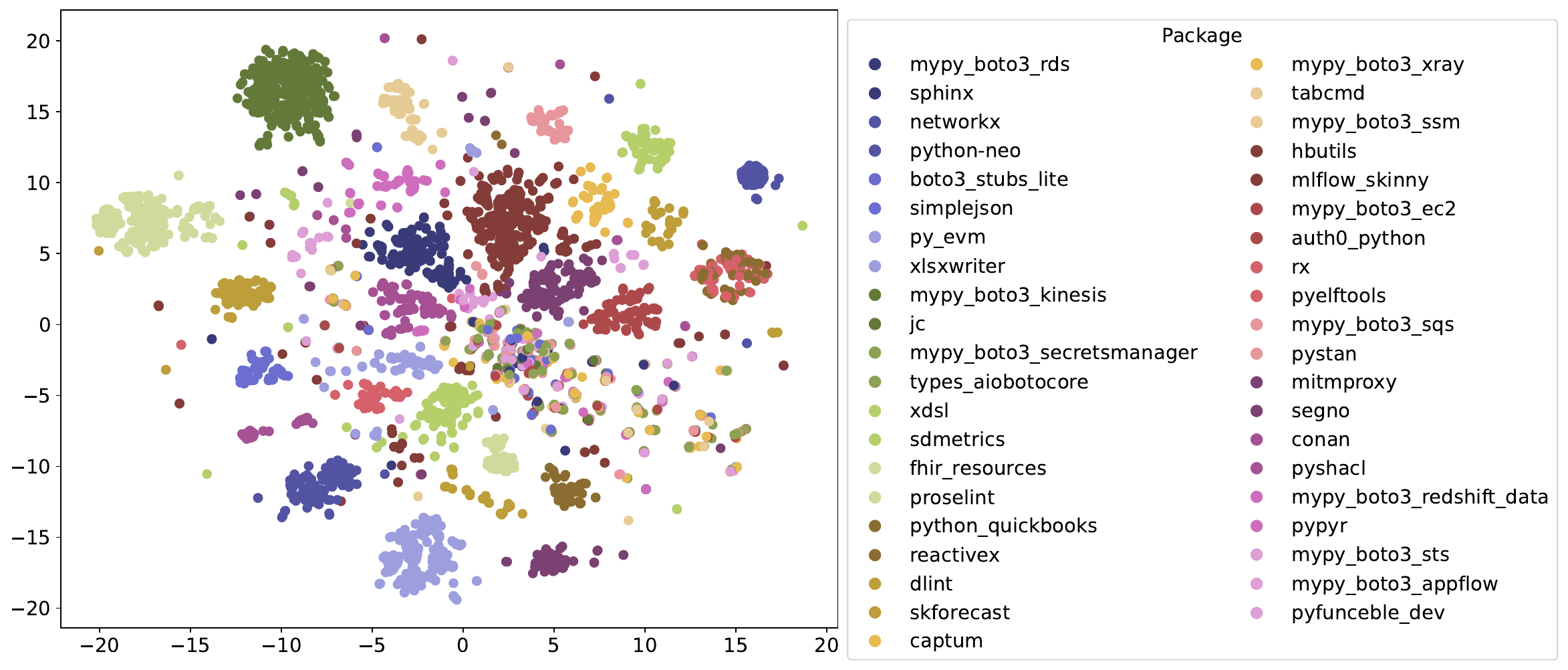}
    \caption{t-SNE visualization of PRD in train set}
    \label{fig:tsne train}
\end{figure}

We assess the diversity of SWE-Dev from two perspectives: sample-level diversity (Figure~\ref{fig:tsne test} and Figure~\ref{fig:tsne train}), and package-level diversity (Figure~\ref{fig: prompt for package classification}).

\textbf{Sample Diversity via t-SNE.}
To visualize the diversity of feature requirements, we perform t-SNE on PRD embeddings generated using OpenAI’s text-embedding-ada-002 model.\footnote{https://platform.openai.com/docs/guides/embeddings} We use 500 test samples and randomly sample 2,000 training samples. Each point represents a PRD, and the color denotes its corresponding package. The resulting distribution reveals rich semantic variation across tasks, even within the same package, highlighting the dataset’s diversity in both content and functionality.

\textbf{Package Category Diversity.}
To analyze the functional diversity of the dataset, we classify packages into high-level categories based on their primary domain (e.g., web development, data science, utilities). The classification is performed using GPT-4o with the prompt provided below (see Figure~\ref{fig: prompt for package classification}). The resulting distribution confirms that SWE-Dev spans a broad spectrum of software domains.

\section{Inference Results}
\label{app:single_infer}

To assess the capabilities and limitations of current LLMs on realistic feature-driven software development tasks, we conduct comprehensive inference-time evaluations on SWE-Dev. We study both single-agent and multi-agent systems, measuring their performance under consistent execution-based evaluation.

\subsection{Single-Agent LLM Performance}
We evaluate 27 state-of-the-art LLMs, including general-purpose chatbot models (e.g., GPT-4o, Claude 3.7) and reasoning models, shown in Table~\ref{tab:overview_of_models}. Models are assessed using Pass@1 and Pass@3 on SWE-Dev's test set. To contextualize benchmark difficulty, we also compare results on HumanEval~\cite{humaneval} and ComplexCodeEval~\cite{complexcodeeval}, using Pass@3 and CodeBLEU respectively. Our findings show that SWE-Dev poses significantly greater challenges than existing benchmarks, with leading models achieving under 30\% Pass@3 on hard tasks (Table~\ref{table: single inference}).

\begin{table*}[h]
\centering
\caption{Evaluation of model performance across benchmarks. This table compares 37 general-purpose and reasoning-focused LLMs on SWE-Dev (Pass@1 and Pass@3 for Easy and Hard splits), ComplexCodeEval (CodeBLEU), and HumanEval (Pass@1 and Pass@3).}
\scalebox{0.5}{
\begin{tabular}{lllccccccc}
\toprule
\multirow{2}{*}{} & \multirow{2}{*}{\textbf{Abbreviated Name}} & \multirow{2}{*}{\textbf{Model}} & \multicolumn{2}{c}{\textbf{SWE-Dev pass@1}} & \multicolumn{2}{c}{\textbf{SWE-Dev pass@3}} & \multirow{2}{*}{\begin{tabular}[c]{@{}c@{}}\textbf{ComplexCodeEval}\\ \textbf{CodeBLEU}\end{tabular}} & \multirow{2}{*}{\begin{tabular}[c]{@{}c@{}}\textbf{HumanEval}\\ \textbf{pass@1}\end{tabular}} & \multirow{2}{*}{\begin{tabular}[c]{@{}c@{}}\textbf{HumanEval}\\ \textbf{pass@3}\end{tabular}} \\
 &  &  & Easy & Hard & Easy & Hard &  &  &  \\ \midrule
\textbf{Chatbot} & Qwen2.5-1.5B & Qwen2.5-1.5B-Instruct & 8.05\% & 1.23\% & 10.76\% & 2.22\% & 29.72\% & 57.20\% & 69.76\% \\
 & Qwen2.5-3B & Qwen2.5-3B-Instruct & 15.93\% & 5.27\% & 21.99\% & 7.47\% & 12.27\% & 62.68\% & 75.00\% \\
 & Qwen2.5-7B & Qwen2.5-7B-Instruct & 25.74\% & 6.68\% & 33.35\% & 7.73\% & 20.00\% & 82.68\% & 87.13\% \\
 & Llama3.1-8B & Llama-3.1-8B-Instruct & 26.43\% & 7.94\% & 33.01\% & 10.24\% & 20.18\% & 68.20\% & 77.87\% \\
 & Qwen3-8B & Qwen3-8B & 34.04\% & 12.09\% & 39.26\% & 13.33\% & 17.47\% & 86.34\% & 89.09\% \\
 & Qwen2.5Coder-14B & Qwen2.5-Coder-14B-Instruct & 39.51\% & 14.82\% & 52.49\% & 18.44\% & 35.52\% & 90.48\% & 92.93\% \\
 & Qwen2.5-14B & Qwen2.5-14B-Instruct & 38.08\% & 13.16\% & 46.32\% & 15.89\% & 19.90\% & 82.56\% & 87.87\% \\
 & DeepseekCoder-V2-Lite & DeepSeek-Coder-V2-Lite-Instruct & 21.53\% & 8.19\% & 29.68\% & 11.33\% & 26.63\% & 80.98\% & 85.18\% \\
 & Qwen3-30B & Qwen3-30B-A3B & 35.84\% & 12.76\% & 39.45\% & 15.20\% & 14.84\% & 89.27\% & 90.30\% \\
 & Phi-4 & Phi-4 & 21.99\% & 5.57\% & 27.89\% & 8.56\% & 33.85\% & 86.46\% & 90.01\% \\
 & Qwen2.5-32B & Qwen2.5-32B-Instruct & 43.64\% & 10.15\% & 51.24\% & 11.69\% & 19.76\% & 88.90\% & 92.44\% \\
 & Qwen2.5-72B & Qwen2.5-72B-Instruct & 49.01\% & 10.62\% & 57.20\% & 12.33\% & 22.15\% & 83.66\% & 86.46\% \\
 & Llama3.3-70B & Llama3.3-70B-Instruct & 33.84\% & 12.85\% & 39.57\% & 14.95\% & 21.29\% & 84.51\% & 88.54\% \\
 & Deepseek-V3 & Deepseek-V3 & 41.95\% & 16.22\% & 56.79\% & 21.62\% & 28.32\% & 90.36\% & 92.92\% \\
 & GPT-4o & GPT-4o & \textbf{54.37\%} & 19.13\% & \textbf{68.70\%} & 21.91\% & 33.38\% & 88.41\% & 92.93\% \\
 & gpt-4o-mini & GPT-4o-mini & 34.47\% & 11.09\% & 41.94\% & 13.84\% & 25.00\% & 85.97\% & 89.00\% \\
 & Claude-3.7-Sonnet & Claude-3.7-Sonnet & 53.09\% & 19.74\% & 56.35\% & 24.25\% & 29.63\% & 93.66\% & 95.36\% \\ \midrule
\textbf{Reasoning} & Claude-3.7-Sonnet(T) & Claude-3.7-Sonnet-thinking & 49.47\% & \textbf{22.51\%} & 56.58\% & \textbf{29.28\%} & \textbf{29.80\%} & 91.22\% & 97.62\% \\
 & Deepseek-R1(Qwen2.5-7B) & Deepseek-R1-distill-Qwen2.5-7B & 6.30\% & 1.29\% & 10.30\% & 1.95\% & 21.05\% & 86.10\% & 93.29\% \\
 & Qwen3-8B(T) & Qwen3-8B-thinking & 19.47\% & 6.36\% & 25.91\% & 9.22\% & 20.98\% & 89.63\% & 91.89\% \\
 & Qwen3-30B(T) & Qwen3-30B-A3B-thinking & 23.63\% & 8.30\% & 31.00\% & 11.60\% & 25.00\% & 93.04\% & 99.57\% \\
 & Deepseek-R1(Qwen2.5-32B) & Deepseek-R1-distill-Qwen2.5-32B & 24.25\% & 9.79\% & 40.53\% & 19.04\% & 27.98\% & 95.17\% & 97.87\% \\
 & Deepseek-R1(Llama-70B) & DeepSeek-R1-distill-Llama-70B & 32.73\% & 8.19\% & 45.72\% & 11.33\% & 25.95\% & 96.95\% & 98.53\% \\
 & Deepseek-R1 & Deepseek-R1-671B & 28.55\% & 12.84\% & 37.62\% & 17.72\% & 34.47\% & \textbf{98.65\%} & \textbf{100\%} \\
 & QwQ-32B & QwQ-32B-Preview & 4.50\% & 0.70\% & 8.90\% & 1.22\% & 24.78\% & 82.31\% & 97.01\% \\
 & Grok-3 & grok-3-beta & \textbf{53.63\%} & 18.97\% & 59.08\% & 22.26\% & 27.96\% & 87.15\% & 89.99\% \\
 & o1 & o1 & 36.36\% & 11.09\% & 43.77\% & 14.27\% & 33.63\% & 97.43\% & 98.78\% \\
 & o3 & o3 & 51.21\% & 21.86\% & 59.05\% & 28.98\% & 26.53\% & 98.04\% & 98.78\% \\ \bottomrule
\end{tabular}}
\label{table: single inference}
\end{table*}

\begin{table*}[h]
\centering
\caption{Information of evaluated LLMs.}
\scalebox{0.75}{
\begin{tabular}{@{}lllcc@{}}
\toprule
\textbf{Model} & \textbf{Size} & \textbf{Release Date} & \textbf{Open} & \textbf{Link} \\ \midrule
Qwen/Qwen2.5-Coder-14B-Instruct & 14B & 2024-11-12 & \checkmark & \href{https://huggingface.co/Qwen/Qwen2.5-Coder-14B-Instruct}{Qwen2.5-Coder-14B-Instruct} \\
deepseek-ai/DeepSeek-Coder-V2-Lite-Instruct & 16B & 2024-06-17 & \checkmark & \href{https://huggingface.co/deepseek-ai/DeepSeek-Coder-V2-Lite-Instruct}{DeepSeek-Coder-V2-Lite-Instruct} \\
microsoft/phi-4 & 14B & 2024-12-12 & \checkmark & \href{https://huggingface.co/microsoft/phi-4}{phi-4} \\
Qwen/Qwen2.5-1.5B-Instruct & 1.5B & 2024-09-19 & \checkmark & \href{https://huggingface.co/Qwen/Qwen2.5-1.5B-Instruct}{Qwen2.5-1.5B-Instruct} \\
Qwen/Qwen2.5-3B-Instruct & 3B & 2024-09-19 & \checkmark & \href{https://huggingface.co/Qwen/Qwen2.5-3B-Instruct}{Qwen2.5-3B-Instruct} \\
Qwen/Qwen2.5-7B-Instruct & 7B & 2024-09-19 & \checkmark & \href{https://huggingface.co/Qwen/Qwen2.5-7B-Instruct}{Qwen2.5-7B-Instruct} \\
meta-llama/Llama-3.1-8B-Instruct & 8B & 2024-07-23 & \checkmark & \href{https://huggingface.co/meta-llama/Llama-3.1-8B-Instruct}{Llama-3.1-8B-Instruct} \\
Qwen/Qwen3-8B & 8B & 2025-04-29 & \checkmark & \href{https://huggingface.co/Qwen/Qwen3-8B}{Qwen3-8B} \\
Qwen/Qwen2.5-14B-Instruct & 14B & 2024-09-19 & \checkmark & \href{https://huggingface.co/Qwen/Qwen2.5-14B-Instruct}{Qwen2.5-14B-Instruct} \\
Qwen/Qwen3-30B-A3B & 30B & 2025-04-29 & \checkmark & \href{https://huggingface.co/Qwen/Qwen3-30B-A3B}{Qwen3-30B-A3B} \\
Qwen/Qwen2.5-32B-Instruct & 32B & 2024-09-19 & \checkmark & \href{https://huggingface.co/Qwen/Qwen2.5-32B-Instruct}{Qwen2.5-32B-Instruct} \\
Qwen/Qwen2.5-72B-Instruct & 72B & 2024-09-19 & \checkmark & \href{https://huggingface.co/Qwen/Qwen2.5-72B-Instruct}{Qwen2.5-72B-Instruct} \\
meta-llama/Llama-3.3-70B-Instruct & 70B & 2024-12-06 & \checkmark & \href{https://huggingface.co/meta-llama/Llama-3.3-70B-Instruct}{Llama-3.3-70B-Instruct} \\
deepseek-ai/DeepSeek-V3 & - & 2024-12-26 & \checkmark & \href{https://huggingface.co/deepseek-ai/DeepSeek-V3}{DeepSeek-V3} \\
gpt-4o (OpenAI) & - & 2024-05-13 & \texttimes & \href{https://openai.com/index/gpt-4o-system-card/}{gpt-4o}/\\
gpt-4o-mini (OpenAI) & - & 2024-07-18 & \texttimes & \href{https://openai.com/index/gpt-4o-mini-advancing-cost-efficient-intelligence/}{gpt-4o-mini} \\
claude-3.7-sonnet (Anthropic) & - & 2025-02-25 & \texttimes & \href{https://www.anthropic.com/claude/sonnet}{claude-3.7-sonnet} \\
grok-3-beta (xAI) & - & 2025-02-19 & \texttimes & \href{https://x.ai/news/grok-3}{grok-3-beta} \\
o1 (OpenAI) & - & 2024-12-05 & \texttimes & \href{https://openai.com/o1/}{o1} \\
o3 (OpenAI) & - & 2025-04-16 & \texttimes & \href{https://openai.com/index/introducing-o3-and-o4-mini/}{o3} \\
deepseek-ai/DeepSeek-R1-Distill-Qwen-7B & 7B & 2025-01-20 & \checkmark & \href{https://huggingface.co/deepseek-ai/DeepSeek-R1-Distill-Qwen-7B}{DeepSeek-R1-Distill-Qwen-7B} \\
deepseek-ai/DeepSeek-R1-Distill-Qwen-32B & 32B & 2025-01-20 & \checkmark & \href{https://huggingface.co/deepseek-ai/DeepSeek-R1-Distill-Qwen-32B}{DeepSeek-R1-Distill-Qwen-32B} \\
Qwen/QwQ-32B-Preview & 32B & 2025-03-06 & \checkmark & \href{https://huggingface.co/Qwen/QwQ-32B-Preview}{QwQ-32B-Preview} \\
deepseek-ai/DeepSeek-R1-Distill-Llama-70B & 70B & 2025-01-20 & \checkmark & \href{https://huggingface.co/deepseek-ai/DeepSeek-R1-Distill-Llama-70B}{DeepSeek-R1-Distill-Llama-70B} \\
deepseek-ai/DeepSeek-R1 & - & 2025-01-20 & \checkmark & \href{https://huggingface.co/deepseek-ai/DeepSeek-R1}{DeepSeek-R1} \\
\bottomrule
\end{tabular}}
\label{tab:overview_of_models}
\end{table*}

\subsection{Multi-Agent System Performance}
We evaluate 10 multi-agent systems (MAS), including both general-purpose MAS (e.g., AgentVerse, LLM Debate) and code-specific designs (e.g., EvoMAC, MapCoder). As detailed in Table~\ref{tab: mas results}, we compare each MAS against a single-agent baseline on execution success (Pass@1), total API call count, and cost-efficiency. Results show that while MAS can outperform single agents on complex tasks, simple strategies (e.g., Self-Refine) often strike a better balance between performance and resource usage than workflow-heavy systems like ChatDev. We visualize the MAS in the following Figure~\ref{fig: mas visualize}.

\begin{figure}
    \centering
    \includegraphics[width=1.0\linewidth]{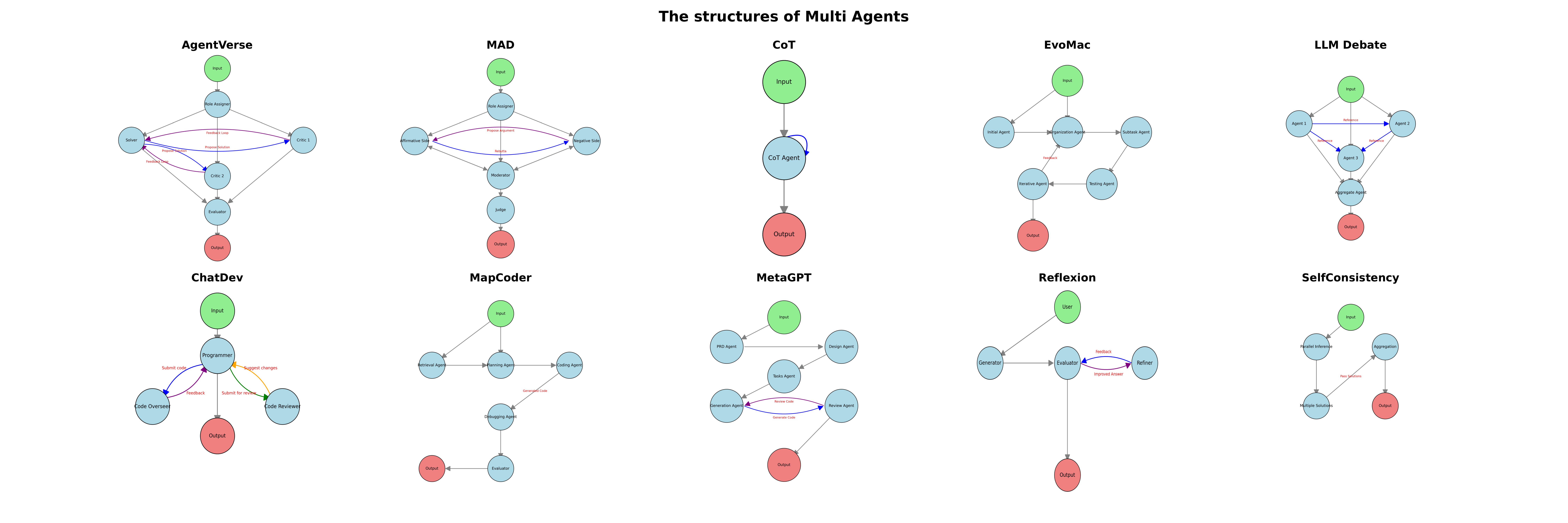}
    \caption{Multiagent system workflow visualization.}
    \label{fig: mas visualize}
\end{figure}

\begin{table*}[]
    \centering
    \captionof{table}{Comparison of general and code-specific MAS on SWE-Dev driven by GPT-4o-mini. \textbf{Bold} highlights the best performance; \underline{underlined} indicates results worse than the single-agent baseline. Most MAS methods outperform the single agent, and simpler general MASs are more effective and efficient than complex coding-specific MASs.}
    \scalebox{0.9}{
\begin{tabular}{l|ccc|ccc}
\toprule
& \multicolumn{3}{c|}{\textbf{Easy}} & \multicolumn{3}{c}{\textbf{Hard}} \\ 
 \multirow{-2}{*}{\textbf{Method}}   & Pass@1 & Calls  & Price(\$) & Pass@1 & Calls  & Price(\$) \\ \midrule
\textbf{Single}                      & 34.47 & 1.00 & 0.75 & 11.09 & 1.00 & 0.97 \\ 
\midrule
\textbf{Reflexion} &   39.77 &
2.12 & 0.83 & 13.32 & 2.18 & 1.35 \\
\textbf{Self Refine}              & \textbf{40.02}   & 5.00     & 5.78  & \textbf{20.03}    & 5.00         & 5.8 \\
\textbf{Self Consistency} & 37.62            & 6.00     & 4.30  & 18.55             & 6.00         & 7.08 \\ 
\textbf{LLM Debate}          & 38.48            & 7.00     & 5.95  & 14.56             & 7.00         & 9.35 \\
\textbf{MAD}                      & \underline{31.50}            & 7.00     & 2.48  & 15.31             & 7.00         & 3.40 \\
 \midrule
\textbf{Agentverse}            & \textbf{38.67}   & 4.52  & 1.40  & 13.42             & 4.83      & 2.90 \\
\textbf{EvoMAC}       & 34.59            & 7.98  & 3.20  & \textbf{13.60}    & 8.30      & 4.65 \\
\textbf{MetaGPT} & \underline{29.56}            & 9.69  & 2.20   & \underline{9.25}             & 10.37     & 4.95 \\
\textbf{MapCoder}                  & \underline{24.55}            & 21.01 & 6.05  & \underline{5.87}              & 23.41     & 10.55 \\

\textbf{ChatDev}                 & 35.13            & 26.61 & 3.53  & 11.70              & 30.87     & 6.10 \\

\bottomrule
\end{tabular}}
\label{tab: mas results}
\end{table*}

\subsection{Agent LLM Results}
\label{app:llm_agent}

\begin{table}[t]
\centering
\caption{Performance of \textbf{tool-augmented LLM agents} on SWE-Dev using the OpenHands framework. We report Pass@1 (\%) on the \textit{Easy} and \textit{Hard} splits. \textbf{Abbreviated names} match those used in the main figures.}
\scalebox{0.88}{
\begin{tabular}{l l cc}
\toprule
\textbf{Abbreviated Name} & \textbf{Model Name} & \textbf{Easy} & \textbf{Hard} \\
\midrule
Claude4.1-Opus(T) & Claude-4.1-Opus-Thinking & 86.30 & 56.44 \\
GPT-5 & GPT-5 & 87.56 & 54.21 \\
Claude4.5(T) & Claude 4.5 Thinking & 84.83 & 53.55 \\
KimiK2(T) & Kimi-K2-Thinking & 83.89 & 50.86 \\
KimiK2 & Kimi-K2-Instruct & 81.82 & 51.05 \\
GLM4.5 & GLM-4.5 & 71.83 & 39.00 \\
Qwen3-235B(T) & qwen3-235b-a22b-thinking-2507 & 43.37 & 19.78 \\
Qwen3-30B(T) & qwen3-30b-a3b-thinking-2507 & 29.68 & 6.11 \\
\bottomrule
\end{tabular}}
\label{tab:agent_results}
\end{table}

\paragraph{Agent Evaluation Setup.}
All agent-based experiments are conducted using the \textbf{OpenHands} \cite{wang2024openhands} framework. We use a unified configuration for all agent LLMs: temperature $=0.6$, maximum agent steps $=100$, execution timeout $=300$ seconds per task, and maximum context length $=64$K tokens. Performance is measured by executing the original SWE-Dev unit tests in the provided environments and reporting Pass@1 on the Easy and Hard splits.

\section{Analysis}
\label{app: swe-dev analysis}

\subsection{Evaluation Metric Validness}
\begin{figure}[t]
        \centering
        \includegraphics[width=0.5\linewidth]{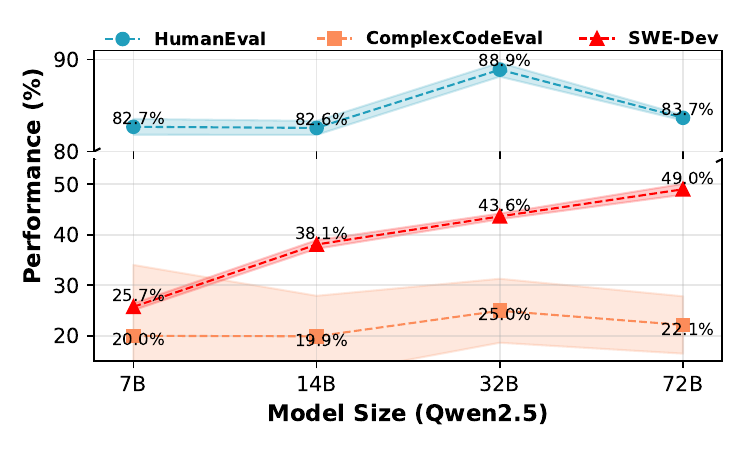}
        \caption{Comparison of benchmarks on various model sizes. SWE-Dev shows clear performance scaling with model size, while HumanEval~\cite{humaneval} fails to distinguish between models. ComplexCodeEval~\cite{complexcodeeval} using CodeBLEU~\cite{codebleu} exhibits high variance, making it less stable for evaluation.}
        \label{fig:other benchmark}

\end{figure}

\textbf{SWE-Dev provides stable and discriminative evaluation of model capabilities.} 
Figure~\ref{fig:other benchmark} compares the performance of Qwen2.5~\cite{qwen2.5} family on SWE-Dev, HumanEval~\cite{humaneval}, and ComplexCodeEval~\cite{complexcodeeval} across three runs. We use Pass@1 for SWE-Dev and HumanEval and ComplexCodeEval for CodeBLEU~\cite{codebleu}. The lines represent the average performance, and the shaded regions show the variance.
We observe that: 
(1) SWE-Dev yields low variance performance and consistent scaling with model size, demonstrating SWE-Dev's stability and reliability in evaluating model capabilities. 
(2) In contrast, HumanEval—despite being stable—is too simple to differentiate models meaningfully.
(3) Meanwhile, ComplexCodeEval shows high variance due to its reliance on similarity-based metrics, CodeBLEU, which limits its reliability for evaluating complex generation tasks.

\subsection{Analysis of PRD Quality}
\label{app: prd_quality analysis}
Our Project Requirement Descriptions (PRDs) are primarily derived from the original docstrings found within the repository source code. To enhance the quality and utility of these PRDs, we employed GPT-4o to refine and improve the original docstrings. To objectively assess this improvement, we recruited two domain experts to evaluate both the original and GPT-4o-enhanced docstrings across 100 randomly selected samples from SWE-Dev.

The experts rated each docstring on three critical dimensions—Clarity, Completeness, and Actionability—using a 0 to 5 scale, where higher scores indicate superior quality. The human evaluation guideline is shown in  Figure~\ref{fig: instruct for prd quality}.

Participants were fully informed about the evaluation process and the nature of the task. The assessment involved only reviewing documentation and posed no ethical or privacy risks, adhering strictly to ethical standards for research involving human subjects. This evaluation provides a rigorous measure of how GPT-4o-refined docstrings enhance PRD quality in SWE-Dev.

\subsection{Analysis on the Underperformance of Reasoning Models}
\label{sec: exp on reasoning}

\textbf{Instruction Following Rate (IFR).} 
Previous experiments have shown that reasoning models perform poorly on SWE-Dev. To investigate the reasons behind this, we analyzed the instruction-following ability of these models.
We measured the percentage of code files that meet the PRD requirements for each model's generated code as a metric of instruction following rate (IFR). The metric is formally defined as: 
\begin{equation*}
    \text{IFR} = \frac{1}{n} \sum_{i=1}^{n} \frac{|\mathcal{G}_i \cap \mathcal{T}_i|}{|\mathcal{T}_i|}
\end{equation*}
where $n$ denotes the total number of tasks, $\mathcal{G}_i$ represents the set of files generated by the model for task $i$, and $\mathcal{T}_i$ denotes the set of ground truth files required by the PRD for task $i$.

\begin{figure}
    \centering
    \includegraphics[width=0.5\linewidth]{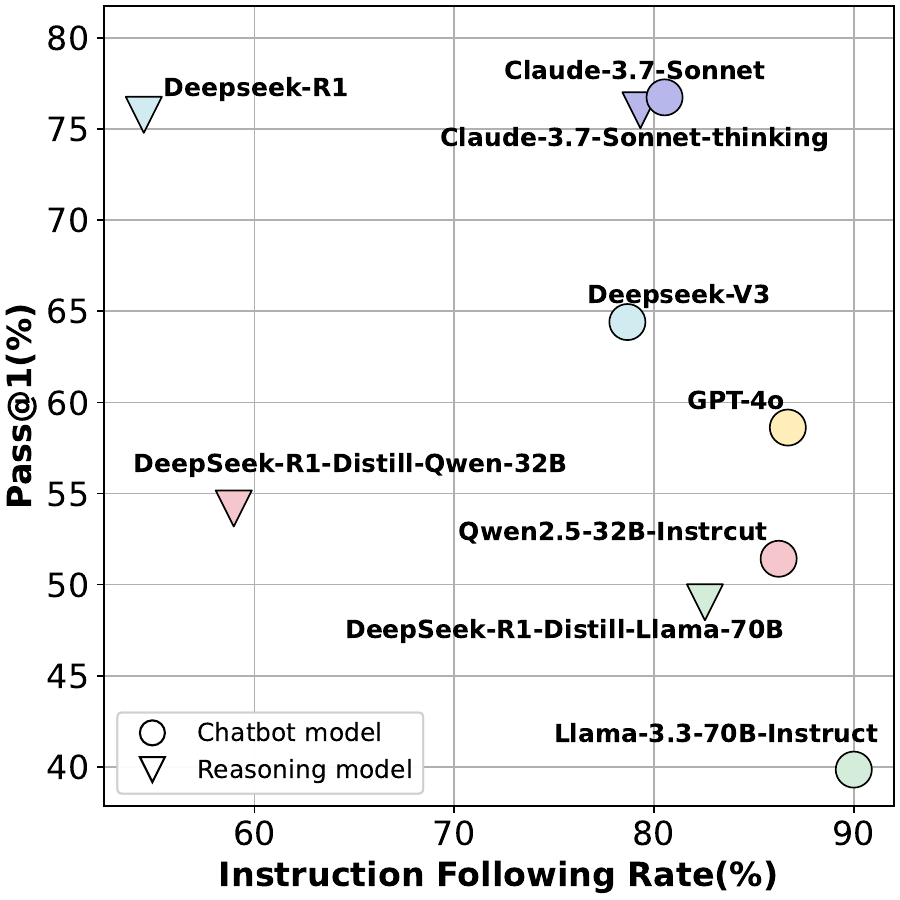}
    \caption{Comparison of reasoning and chatbot LLMs' IFR and performance on SWE-Dev. Reasoning models tend to outperform chatbots when they fully follow instructions, though their overall IFR is lower.}
    \label{fig:abl_reasoning}
\end{figure}

To further explore this, we compared reasoning models with their chatbot counterparts by evaluating their instruction following rate. Specifically, in Figure~\ref{fig:abl_reasoning}, the x-axis represents the instruction following rate, and the y-axis shows the performance of both reasoning models and their chatbot counterparts on tasks where their instruction following rate is 100\%.

As shown in the figure, we see that: 
(i) Reasoning models generally have a lower instruction-following rate compared to their chatbot version, which explains why they underperform when handling multiple tasks simultaneously. Reasoning models tend to struggle with tasks on SWE-Dev that involve performing several steps in a single call, resulting in poorer performance overall. 
(ii) However, on tasks where both reasoning models and their chatbot versions have an instruction-following rate of 100\%, reasoning models typically outperform the chatbots. This indicates the potential of reasoning models when they can fully adhere to instructions. 
(iii) Claude 3.7-Sonnet is an exception to this trend, as both its reasoning and chatbot versions exhibit similar instruction-following rates and performance, which contributes to Claude's superior results.

\subsection{Analysis of MAS}
\label{subsec:mas_communication_analysis}

While multi-agent systems (MAS) demonstrate performance gains over single-agent baselines on SWE-Dev, their behaviors vary significantly across designs. To better understand these differences, we analyze MAS from a communication and coordination perspective. Specifically, we examine how agent structure and interaction overhead relate to execution success on repository-level feature development tasks.

\begin{table*}[t]
\centering
\small
\setlength{\tabcolsep}{5pt}
\begin{tabular}{lcccc}
\toprule
\textbf{MAS System} & \textbf{Pass@1 (\%)} & \textbf{Avg. LLM Calls} & \textbf{\# Roles} & \textbf{Avg. Agent-to-Agent Msgs} \\
\midrule
Reflexion & 39.77 & 2.12 & 3 & 10.77 \\
Self-Refine & \textbf{40.02} & 5.00 & 2 & 26.79 \\
Self-Consistency & 37.62 & 6.00 & 2 & 33.01 \\
LLM Debate & 38.48 & 7.00 & 2 & 76.35 \\
MAD & 31.50 & 7.00 & 4 & 2.36 \\
AgentVerse & 38.67 & 4.52 & 4 & 8.83 \\
MetaGPT & 29.56 & 9.69 & 6 & 24.26 \\
MapCoder & 24.55 & 21.01 & 5 & 63.03 \\
\bottomrule
\end{tabular}
\vspace{2pt}
\caption{\textbf{Communication and coordination statistics of MAS on SWE-Dev.}
All systems are driven by the base model (GPT-4o-mini).
We report Pass@1, average number of LLM calls per task, number of predefined agent roles,
and the average number of agent-to-agent messages.}
\label{tab:mas_comm_analysis}
\end{table*}

Table~\ref{tab:mas_comm_analysis} reveals two consistent trends.
First, simple and low-overhead MAS achieve the best performance--efficiency trade-off.
General-purpose methods such as Reflexion~\cite{reflexion} and Self-Refine~\cite{self_refine} use only two to three roles and require relatively few model calls.
Despite their simplicity, they reach the highest Pass@1 scores.
For example, Reflexion achieves 39.77\% Pass@1 with only 2.12 LLM calls and 10.77 agent-to-agent messages on average.
These systems rely on iterative self-refinement rather than explicit division of labor.
As a result, the global task context remains concentrated within a small number of agents, reducing coordination cost.

Second, workflow-heavy and code-specific MAS suffer from coordination overhead and context fragmentation.
Systems such as MetaGPT~\cite{hong2024metagptmetaprogrammingmultiagent} and MapCoder~\cite{islam2024mapcodermultiagentcodegeneration} introduce more roles, more calls, and substantially more inter-agent communication.
However, this increased interaction does not translate into better execution success.
MapCoder, for instance, uses over 21 LLM calls and exchanges more than 60 agent-to-agent messages on average,
yet achieves the lowest Pass@1 among all compared methods.
Each additional handoff between agents introduces opportunities for partial context loss and error propagation.
This effect is particularly harmful on SWE-Dev tasks, which already involve long code contexts and multi-file dependencies.

Overall, these results suggest that, for repository-level feature development, maintaining a stable and unified reasoning context is more important than fine-grained role specialization.
Simple self-refinement style MAS exhibit an emergent robustness by minimizing communication overhead while preserving global task coherence.
In contrast, heavily engineered multi-role workflows tend to accumulate coordination cost without corresponding performance gains.

\subsection{Error Analysis}
\begin{figure}[h]
    \centering
    \includegraphics[width=1.0\linewidth]{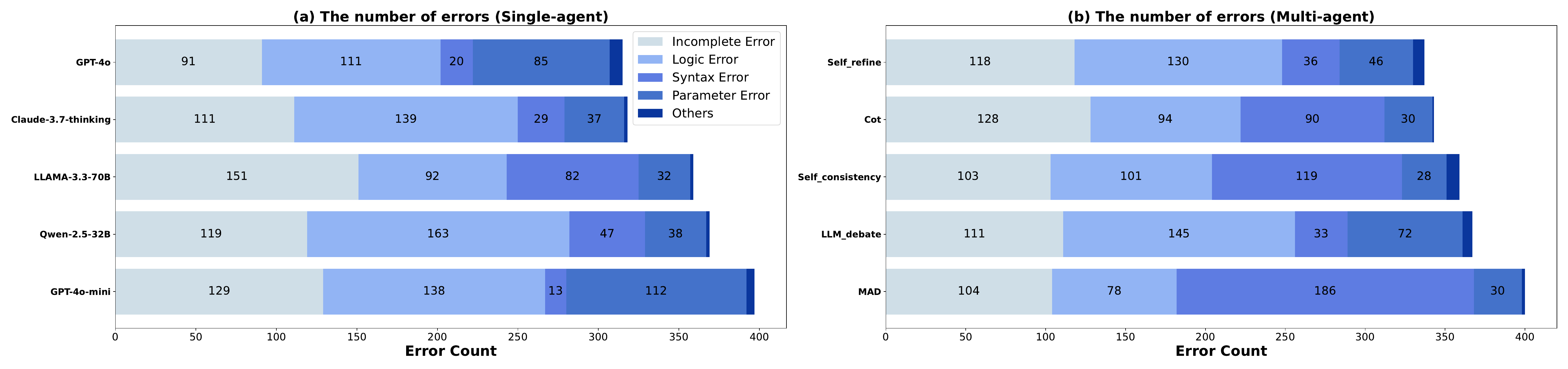}
    \caption{Failure case distribution of Single and Multi-agent.}
    \label{fig:error classification}
\end{figure}

Figure~\ref{fig:error classification} presents the distribution of failure types for both single-agent and multi-agent systems on SWE-Dev. We sample 500 samples for error analysis and categorize errors into five types: Incomplete, Logic, Syntax, Parameter, and Others, see error classification prompt in Figure~\ref{fig: prompt for error classification}. Across both agent types, the most prevalent error is the Incomplete Error, where models fail to implement PRD-required functions—indicating persistent challenges in task decomposition and execution coverage.

For single-agent models, Logic Errors are the second most common, followed by Parameter Errors and Syntax Errors. Interestingly, GPT-4o and Claude-3.7 show relatively fewer Syntax Errors, suggesting better adherence to Python syntax, while smaller models like GPT-4o-mini show higher incidence of both Syntax and Parameter issues, reflecting their limited reasoning capacity and weaker control over function signatures.

In contrast, multi-agent systems exhibit a different pattern. While they reduce Incomplete Errors to some extent, they often incur higher Logic or Syntax Errors—especially in methods like MAD and Self-consistency—suggesting that while agents may cover more PRD content, coordination breakdown or hallucinated reasoning steps can introduce new failure modes.

Overall, the analysis highlights the need for improved function selection, robust reasoning alignment, and stronger control over generation structure—especially in collaborative multi-agent settings.

\subsection{Limitation and Future Work}
\label{sec; limitation}

\paragraph{Language Scope.}
SWE-Dev currently targets Python, which, while widely used, does not reflect the full diversity of real-world programming languages. A natural extension is to support other major languages such as Java, JavaScript, and C++, enabling broader evaluation and enhancing generality.

\paragraph{Training Exploration.}
Our training experiments focus on standard techniques—SFT, RL, and role-wise MAS training—which yield modest gains. Future work could explore stronger RL~\cite{grpo,dapo,reinforce++}, dynamic agent coordination~\cite{masgpt}, and curriculum learning~\cite{bengio2009curriculum}. Notably, SWE-Dev offers fine-grained complexity signals via call trees that can guide complexity-aware training.

\subsection{Broader impacts}
\label{sec: impacts}
SWE-Dev is the first dataset tailored for autonomous feature-driven software development, addressing the gap between current automated coding and real-world software engineering demands.
By providing large-scale, realistic tasks based on real repositories with executable tests, it enables rigorous and reliable evaluation of automated AI coding systems. 
SWE-Dev promotes the creation of more capable methods for complex software, driving innovation that can lower development costs and enhance software quality industry-wide.

\section{Detailed Benchmark Construction}

\subsection{Call tree generation}
To accurately localize the implementation logic associated with each test case, we construct a call tree that captures the dynamic execution path from the test to the relevant source functions. This tree serves as the foundation for identifying the core feature logic and determining task complexity.

\begin{figure}
    \centering
    \includegraphics[width=0.5\linewidth]{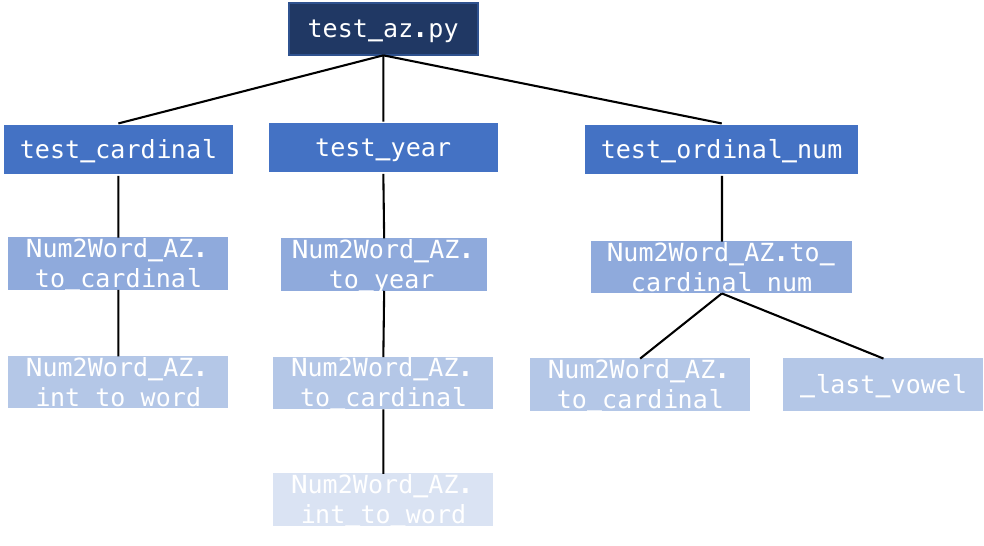}
    \caption{Example of a generated call tree for \texttt{test\_az.py}.}
    \label{fig:example_call_tree}
\end{figure}

Figure~\ref{fig:example_call_tree} shows a generated call tree for the file \texttt{test\_az.py}, which contains multiple test functions such as \texttt{test\_cardinal}, \texttt{test\_year}, and \texttt{test\_ordinal\_num}. Each test function serves as a root for its own call path, triggering downstream functions like \texttt{Num2Word\_AZ.to\_c} \texttt{ardinal} and \texttt{Num2Word\_AZ.int\_to\_word}. This tree structure reveals the multi-level and cross-functional logic activated during test execution, illustrating how test files connect to multiple feature implementations across the codebase.

We use the call tree in two key ways:
\begin{enumerate}
    \item To select target functions for masking during task generation, enabling controllable task complexity.
    \item To trace which source files and logic a model must understand to solve the task, supporting fine-grained evaluation and curriculum learning.
\end{enumerate}

\subsection{Docstring Augmentation Prompt} 
To ensure high-quality task specifications, we augment original function-level docstrings using GPT-4o. Figure~\ref{fig: prompt for docstring augmentation} shows the prompt we use to generate concise, informative docstrings conditioned on the full code context.

\subsection{Masking Procedure and Reproducibility}

Our dataset construction does not rely on any task-complexity thresholds such as call-tree depth or node count. Instead, we apply a fixed set of masking rules uniformly across all repositories. Specifically, for each task we derive the dynamic call chain exercised by the upstream test and construct a masked input using one of three modes: (i) masking depth-1 target functions, (ii) masking depth-2 dependent functions, or (iii) masking the entire call chain for call-tree depth $\geq$ 2. These masking modes are predetermined and do not involve any further filtering or parameter tuning.

\section{Extended Related Work}
\subsection{Multi-agent system}
For complex SE tasks that strain the context handling of single agents, Multi-Agent Systems (MAS) utilizing collaborating LLMs are an emerging research avenue. Existing frameworks like MetaGPT, ChatDev, and AgentVerse often rely on predefined agent roles and fixed interaction protocols. While effective on specific tasks, their hand-crafted structure limits generalization. Recent research explores trainable MAS, aiming for agents that dynamically adapt their organization or communication strategies. However, empirical studies of such adaptive MAS are largely constrained by benchmark complexity; evaluations are often confined to small-scale or synthetic tasks due to the lack of benchmarks providing complex interaction scenarios and reliable execution feedback. SWE-Dev's scale, complexity, and provision of executable feedback (via unit tests) establish it as the first testbed capable of supporting the training and evaluation of dynamic MAS on realistic, multi-file feature development scenarios.

\section{Experiment Settings}
\label{app: settings}
\subsection{Inference}
\label{app:inference}

\textbf{LLMs.} 
We evaluate 17 chatbot LLMs with different model size, including Qwen2.5-Instruct models 1.5B/3B/7B/14B/32B/72B~\cite{qwen2025qwen25technicalreport}, Qwen3 models 8B/30B-A3B~\cite{qwen3}, Llama 3.1-8B/3.3-70B-Instruct~\cite{grattafiori2024llama3herdmodels}, Phi 4~\cite{abdin2024phi4technicalreport}, Claude-3.7-Sonnet~\cite{claude3.7}, Deepseek-V3~\cite{deepseekai2025deepseekv3technicalreport}, GPT-4o~\cite{gpt4o}, Deepseek-Coder-V2-Lite-Instruct~\cite{deepseekai2024deepseekcoderv2breakingbarrierclosedsource}, Qwen2.5-Coder-14B-Instruct~\cite{hui2024qwen25codertechnicalreport}. 
Additionally, We extend the evaluation to reasoning models, including Deepseek-R1-distill models (Qwen2.5 7B/32B, Llama-70B)~\cite{deepseek}, Qwen3 8B/30B-A3B (thinking)~\cite{qwen3}, QwQ-32B-Preview, Deepseek-R1~\cite{deepseek}, OpenAI-o1~\cite{openai2024openaio1card}, Claude-3.7-Sonnet-thinking~\cite{claude3.7}, and Grok-3-Beta~\cite{grok3}. 

\textbf{Multi-Agent Systems.} 
\label{app: MAS}
To provide a more comprehensive evaluation of SWE-Dev, we expand our study to include multi-agent systems (MAS) built on LLMs. Prior research has demonstrated that MAS can enhance performance on tasks requiring multi-step reasoning and coordination~\cite{masgpt,hong2024metagptmetaprogrammingmultiagent,qian2024chatdevcommunicativeagentssoftware}. 
In our experiments, all MAS are implemented using GPT-4o-mini~\cite{gpt4omini} as the underlying model to ensure consistency across methods. And for fair comparison, we utilize MASLab~\cite{ye2025maslabunifiedcomprehensivecodebase}, a unified framework integrating multiple MAS implementations. 
We evaluate coordination-based MAS such as LLM Debate~\cite{llm_debate}, Self Refine~\cite{self_refine}, Multi-Agent Debate (MAD)~\cite{mad}, and Self Consistency~\cite{self_consistency} that feature relatively simple agent interaction strategies. 
We further include structured, workflow-oriented MAS designed for code generation, including Agentverse~\cite{chen2023agentversefacilitatingmultiagentcollaboration}, MetaGPT~\cite{hong2024metagptmetaprogrammingmultiagent}, ChatDev~\cite{qian2024chatdevcommunicativeagentssoftware}, MapCoder~\cite{islam2024mapcodermultiagentcodegeneration}, and EvoMAC~\cite{evomac}.

\subsection{Training}
\label{app: train}
\subsection{Single-Agent Supervised Fine-tuning}
\label{app: train sft}
We fine-tune the model using LoRA, applying low-rank adaptations (rank $r = 16$, scaling $\alpha = 16$, $dropout = 0.05$) to the query, key, value, and output projection matrices of each attention sublayer. Training is performed with a learning rate of $6 \times 10^{-4}$ and a batch size of 32 sequences per gradient step, for up to 4 epochs. Checkpoints are saved every 50 steps, and the best model is selected based on validation loss over a held-out set of 100 examples. Fine-tuning is initialized from Qwen2.5-7B-Instruct and completed within 20 hours using 8 NVIDIA A100 GPUs. We leverage DeepSpeed Ulysses and Flash Attention to support efficient training with long input contexts.

\subsection{Single-Agent Reinforcement Learning}
\label{app: train rl}

For reinforcement learning (RL) training, we sampled 2k instances from SWE-Dev to balance computational feasibility and the ability to capture RL benefits. Specifically, we used Proximal Policy Optimization (PPO)~\cite{ppo} and Direct Preference Optimization (DPO)~\cite{dpo} for training the Qwen2.5-7B-Instruct using 8 NVIDIA A100 GPUs. All RL experiments are run on a single node with 8$\times$A100 GPUs, using VERL PPO with a vLLM-based rollout backend.

\textbf{PPO}: 
The training was conducted using a batch size of 256 and trained for 5 epochs, with a learning rate of $1 \times 10^{-6}$ for the actor and $1 \times 10^{-5}$ for the critic. We set the KL coefficient to 0.001. The training was set to save checkpoints every 10 steps. We used a maximum prompt length of 8192 tokens and set a micro-batch size of 32. The maximum response length is 2048 tokens. The reward for PPO is calculated based on the pass rate of the test cases. Concretely, for each rollout solution, the reward is defined as the pass rate, i.e., the number of test cases successfully passed by the solution divided by the total number of test cases available for that task. During RL training, we do not employ any external agent framework: the language model directly receives the problem requirement description (PRD) together with the relevant code context as input and generates a single rollout solution.

\textbf{DPO}: 
For DPO training, we applied LoRA with a rank of 64, scaling factor $\alpha = 128$, and dropout set to 0. The preference loss function (\texttt{pref\_loss}) was set to sigmoid, which is commonly used in DPO for preference-based optimization. Training was performed for 5 epochs, using a batch size of 8 and a learning rate of $1 \times 10^{-5}$.

    
    
    
    

For a fair comparison with SFT in~\S\ref{sec: single rl}, we used the same 2k training samples for both SFT and RL. The details for SFT training are outlined in Appendix~\ref{app: train sft}.

These methods allow us to assess the impact of RL on model performance using the SWE-Dev dataset while maintaining efficient training..

\begin{figure}
    \centering
    \includegraphics[width=0.8\linewidth]{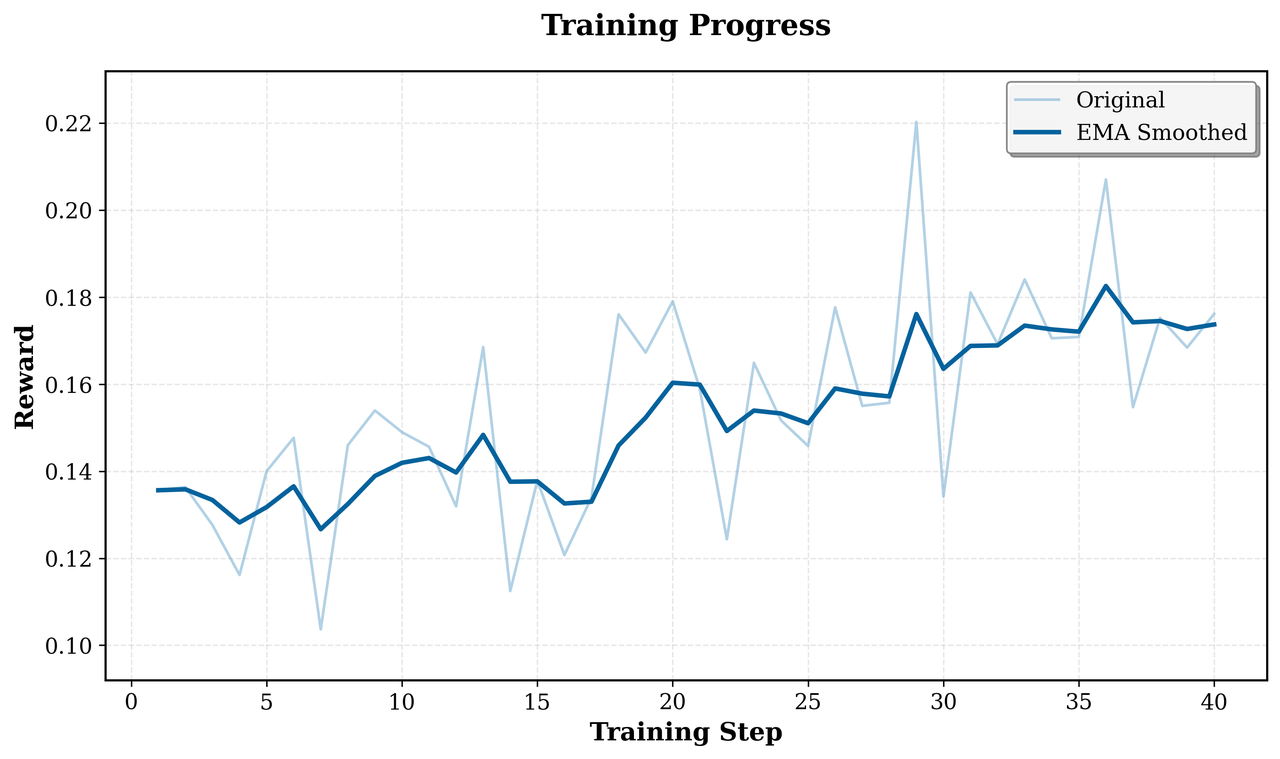}
    \caption{The PPO training progress of Qwen2.5-7B-Instrut on the SWE-Dev with 40 steps and 2048 max token length. The light curve shows the raw reward, while the dark curve applies EMA smoothing.}
    \label{fig:placeholder}
\end{figure}

\begin{figure}
    \centering
    \includegraphics[width=0.8\linewidth]{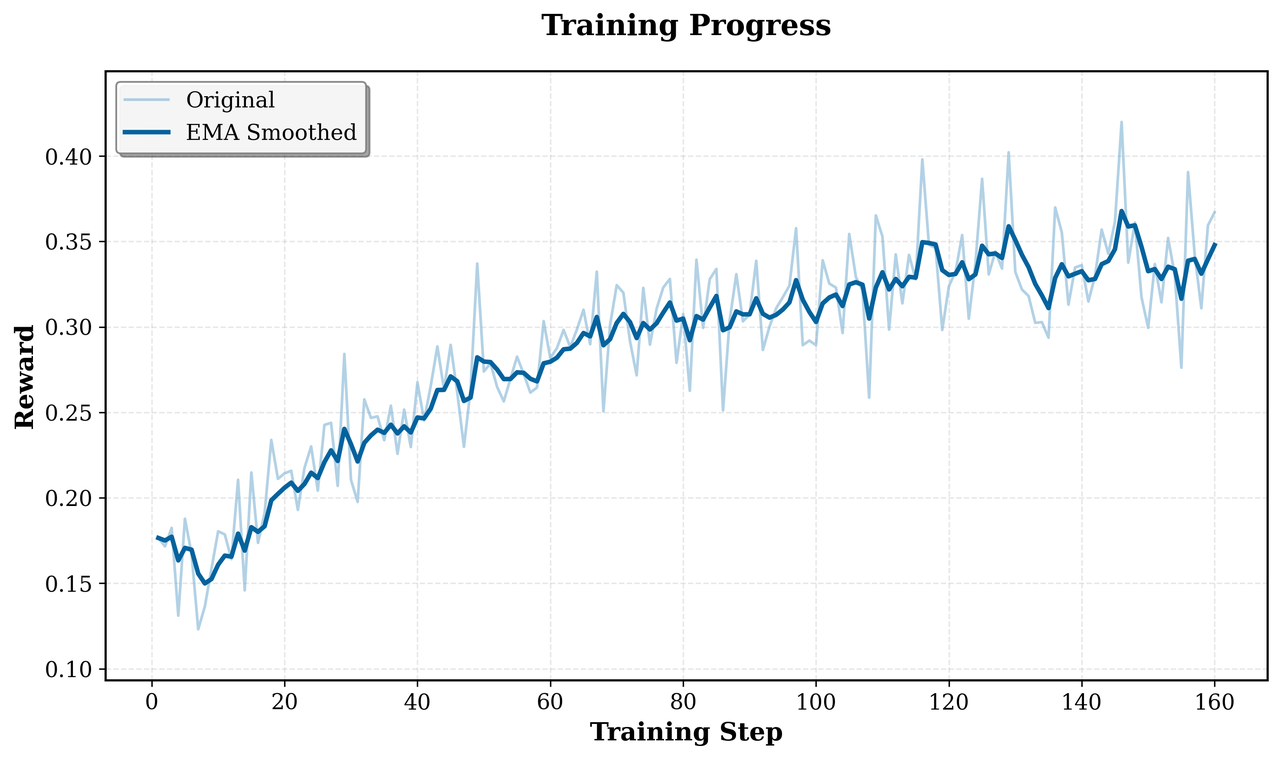}
    \caption{The PPO training progress of Qwen2.5-7B-Instrut on the SWE-Dev with 160 steps and 8192 max token length. The light curve shows the raw reward, while the dark curve applies EMA smoothing.}
    \label{fig:placeholder}
\end{figure}

\begin{figure}
    \centering
    \includegraphics[width=0.8\linewidth]{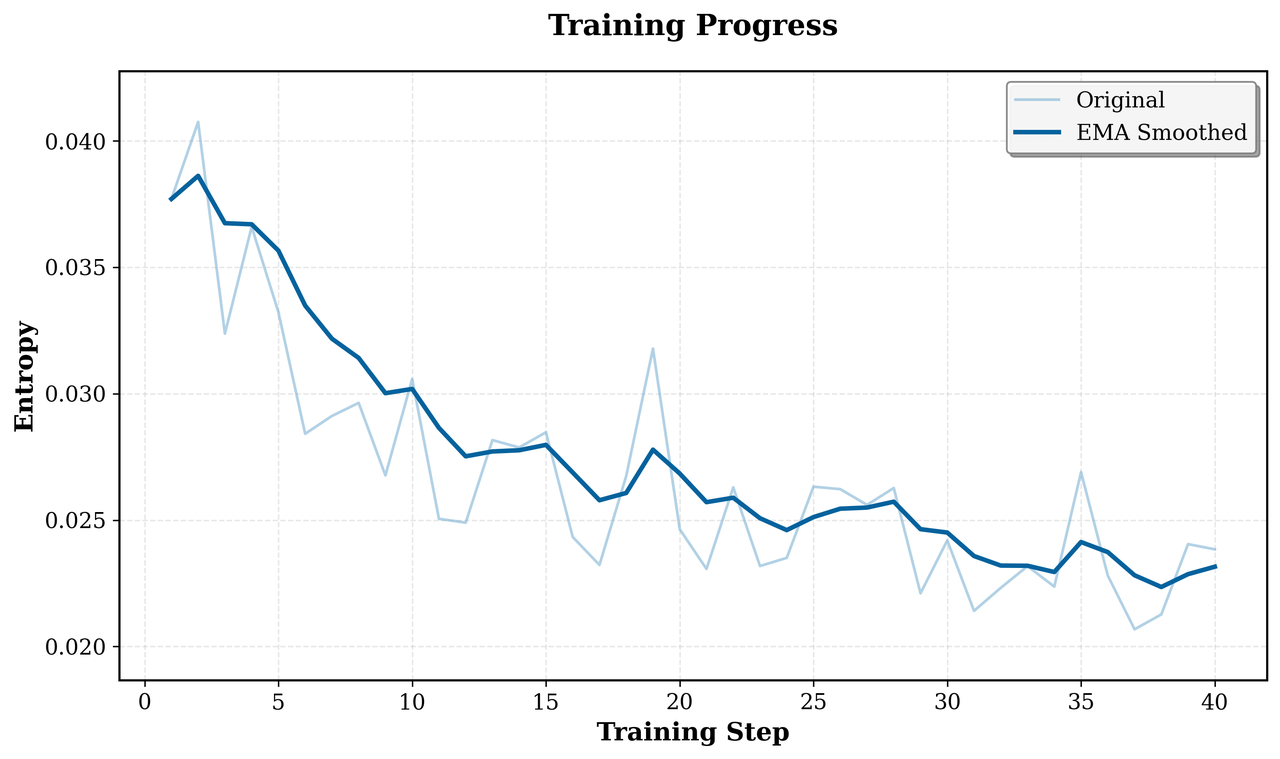}
    \caption{Entropy trajectory during PPO training (40 updates). The light curve shows the raw policy entropy, and the dark curve shows the EMA-smoothed trend. Entropy decreases steadily over training, indicating reduced exploration and a progressively sharper output distribution—consistent with the observed contraction in diversity under RL.}
    \label{fig:placeholder}
\end{figure}
\subsection{Multi-Agent Supervised Fine-tuning}

\begin{figure}
    \centering
    \includegraphics[width=\linewidth]{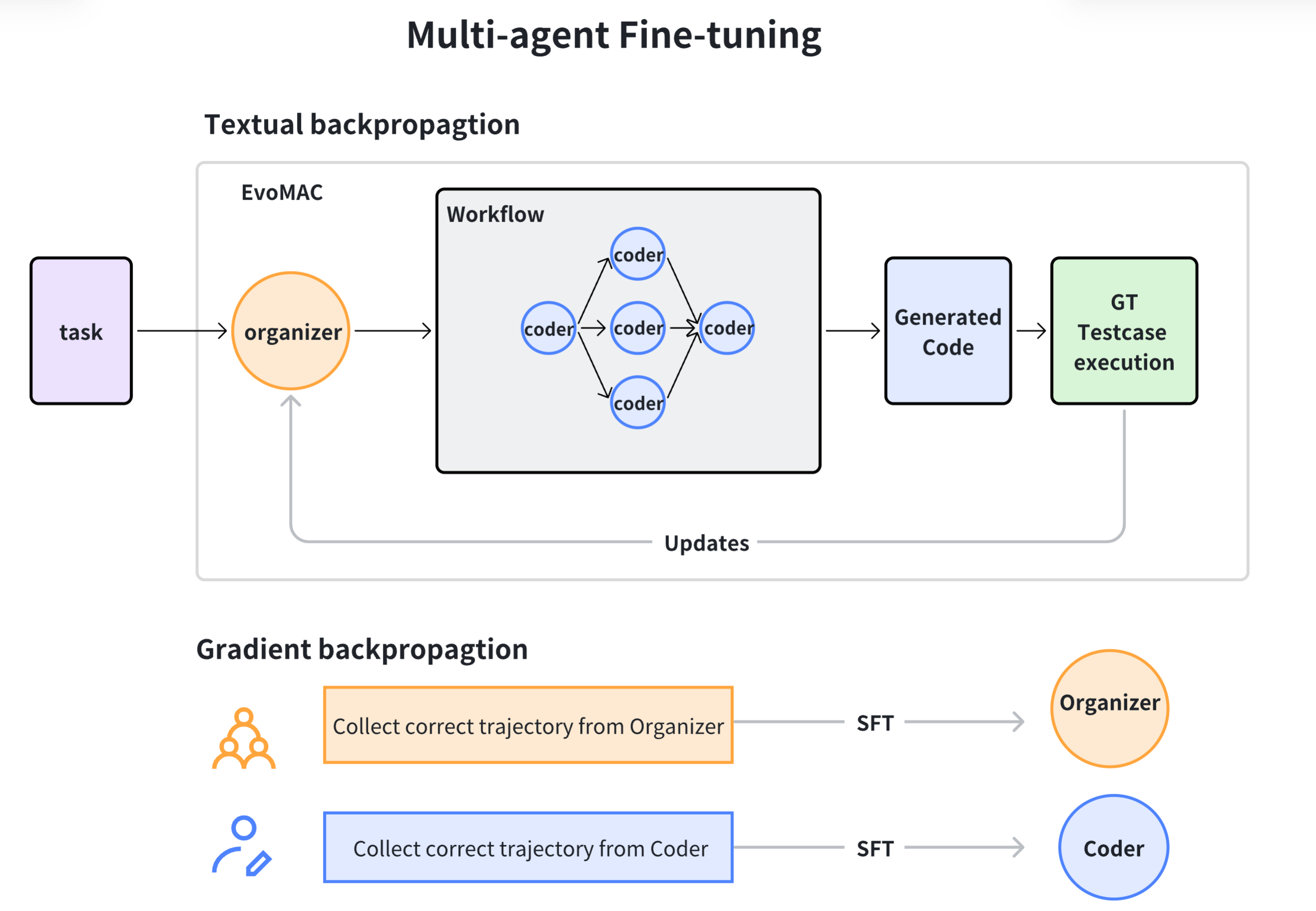}
    \caption{Overview of Multi-agent Fine-tuning in EvoMAC. This framework leverages both textual and gradient-based supervision to improve multi-agent collaboration. During inference, an organizer assigns roles and coordinates a team of coders to generate code, which is then validated using ground truth test cases. Successful execution trajectories are collected and used to fine-tune both the organizer and coders individually via supervised learning, enabling role-specific optimization for complex software development tasks.}
    \label{fig: evomac}
\end{figure}

In our multi-agent fine-tuning experiments, we utilize a simplified version of EvoMAC~\cite{evomac}, retaining only two core roles: \textbf{Organizer} and \textbf{Coder}, see Figure~\ref{fig: evomac}. The fine-tuning process follows an iterative workflow. Initially, the \textbf{Organizer} processes the Project Requirement Description (PRD) and breaks it down into clearly defined subtasks or instructions. Subsequently, the \textbf{Coder} generates corresponding code implementations for these subtasks. The generated code is then evaluated using the provided ground truth (GT) test cases. Feedback from these evaluations informs subsequent iterations, enabling iterative refinement of both the task decomposition by the Organizer and the code generation by the Coder. Fig.~\ref{fig: prompt for EvoMAC coding agent} and Fig.~\ref{fig: prompt for EvoMAC organizing agent} respectively show the prompts used for the Coder and Organizer.

\textbf{Rejection Sampling Procedure.}

To effectively leverage the feedback from GT test cases, we employ rejection sampling—a method widely adopted in reinforcement learning and language model fine-tuning~\cite{zhou2024archertraininglanguagemodel,snell2023offlinerlnaturallanguage}. The detailed procedure is as follows:

\begin{enumerate}
\item \textbf{Iterative Reasoning with EvoMAC}: For each training instance, EvoMAC executes multiple rounds of reasoning. In each iteration, the generated code from the Coder is tested against the GT test cases to compute its performance.

\item \textbf{Selection of High-quality Trajectories}: Trajectories that show improved performance over previous iterations (as indicated by an increased pass rate on GT test cases) are selectively retained. Conversely, trajectories that do not demonstrate progress or degrade in performance are discarded. This ensures that only beneficial and constructive data is used for fine-tuning.

\item \textbf{Role-wise Fine-tuning}: The retained high-quality trajectories are utilized to separately fine-tune the Organizer and the Coder. Specifically, the Organizer is trained to better structure and decompose tasks from PRDs, while the Coder is refined to enhance code generation capabilities for defined subtasks. This role-specific fine-tuning promotes specialization and improves overall performance.

\end{enumerate}

As shown in~\S\ref{sec: multi train}, through this simplified EvoMAC and structured rejection sampling approach, our multi-agent fine-tuning effectively enhances the capabilities of each agent, contributing to significant performance gains on SWE-Dev.

\section{Licensing}
\label{app: licensing}

All codebases and data used in this work are sourced from publicly available GitHub repositories. We have ensured compliance with the corresponding licenses of these repositories, respecting all terms of use and attribution requirements.

\section{Prompts}
This section includes all prompts used in the generation, evaluation and analysis process.

\label{app: prompt}
\begin{figure*}
\begin{tcolorbox}[title=Single LLM inference prompt]
\lstset{
    basicstyle=\normalfont\sffamily\footnotesize,
    breaklines=true,
    frame=none,
    columns=fullflexible,
}
\begin{lstlisting}[breaklines=true, basicstyle=\ttfamily\small]
# AIM: You need to assist me with a Python package feature development task. I will provide a Product Requirements Document (PRD) that details the functionality and lists the empty functions that need to be implemented across different file paths. I will also provide the complete "Code Context" of all files mentioned in the PRD.

Your task is to implement ONLY the empty functions described in the PRD while preserving ALL OTHER CODE in the provided files exactly as is. This is absolutely critical - you must keep all imports, class definitions, functions, comments, and other code that is not explicitly mentioned in the PRD for implementation.

When implementing the functions:
1. Carefully identify which functions from the PRD need implementation.Implement them based on the docstrings and specifications in the PRD
2. Do not add any new "import" statements unless absolutely necessary
3. Do not modify ANY existing code structure, only implement the empty functions

For each file mentioned in the PRD, you MUST output the COMPLETE file code with your implementations inserted. Your output format must follow this exact pattern:

# OUTPUT FORMAT FOR EACH FILE:
@ [relative path/filename]
```python
[COMPLETE file code including ALL original code plus your implementations]
```
@ [relative path/filename]
```python
[COMPLETE file code including ALL original code plus your implementations]
```

IMPORTANT: Make sure your output includes EVERY function, class, import statement, and comment from the original code context. The only difference should be that the empty functions specified in the PRD are now implemented.

# PRD:
{PRD}

# Code Context:
{code_snippet}
\end{lstlisting}
\end{tcolorbox}
\caption{Single LLM Inference Prompt.}
\label{fig: single agent inference prompt}
\end{figure*}

\begin{figure*}
\begin{tcolorbox}[title=EvoMAC Coding Agent Prompt]
\lstset{
    basicstyle=\normalfont\sffamily\footnotesize,
    breaklines=true,
    frame=none,
    columns=fullflexible,
}
\begin{lstlisting}[breaklines=true, basicstyle=\ttfamily\small]
You are Programmer. we are both working at ChatDev. We share a common interest in collaborating to successfully complete a task assigned by a new customer.

You can write/create computer software or applications by providing a specific programming language to the computer. You have extensive computing and coding experience in many varieties of programming languages and platforms, such as Python, Java, C, C++, HTML, CSS, JavaScript, XML, SQL, PHP, etc,.

Here is a new customer's task: {task}.

To complete the task, you must write a response that appropriately solves the requested instruction based on your expertise and customer's needs.

According to the new user's task and you should concentrate on accomplishing the following subtask and pay no heed to any other requirements within the task.

Subtask: 
{subtask}.

Programming Language: python,

Codes:
{codes}
\end{lstlisting}
\end{tcolorbox}
\caption{Role Prompt for EvoMAC Coding Agent.}
\label{fig: prompt for EvoMAC coding agent}
\end{figure*}

\begin{figure*}
\begin{tcolorbox}[title=EvoMAC Organizing Agent Prompt]
\lstset{
    basicstyle=\normalfont\sffamily\footnotesize,
    breaklines=true,
    frame=none,
    columns=fullflexible,
}
\begin{lstlisting}[breaklines=true, basicstyle=\ttfamily\small]
As the Leader of a coding team, you should analyze and break down the problem into several specific subtasks and assign a clear goal to each subtask. Ensure each subtask is extremely detailed, with a clear description and goal, including the corresponding PRD statement where necessary.

The workflow should be divided into minimal, executable subtasks, each involving one method implementation. The target_code should only contain the relative paths and function names for the specific code that is required for that subtask.

Each subtask should be assigned a unique task_id, and the description should reflect the exact requirements of the PRD corresponding to that method or task. The target_code should be precise, containing only the specific Python code (relative path and method/function name) that corresponds to the subtask's scope.

The format should strictly follow the JSON structure below:

```json
[
    {
        "task_id":"1",
        "description":"Task Description",
        "target_code": [
            "relative_python_path:function_name", 
            "relative_python_path:class_name.method_name"
        ]
    }
]
```

Use the backticks for your workflow only.

Note: 

(1) Each subtask should be self-contained and represent one method's implementation.

(2) The `description` should be based on specific statements from the PRD, and it must explain what the subtask is aiming to achieve.

(3) The `target_code` should only reference the code paths and function names for the methods to be implemented for the subtask.

(4) The number of subtasks should not exceed 5. Some tasks might combine multiple smaller functions if needed to fit within the limit.

(5) Each subtask is handled independently by different agents, so the description should be thorough, ensuring clarity even without the full context of the PRD.
\end{lstlisting}
\end{tcolorbox}
\caption{Role Prompt for EvoMAC Organizing Agent.}
\label{fig: prompt for EvoMAC organizing agent}
\end{figure*}

\begin{figure*}
\begin{tcolorbox}[title=Error Classification Prompt]
\lstset{
    basicstyle=\normalfont\sffamily\footnotesize,
    breaklines=true,
    frame=none,
    columns=fullflexible,
}
\begin{lstlisting}[breaklines=true, basicstyle=\ttfamily\small]
You are an error classification expert. Based on the provided PRD, LLM-generated code, and error message, your task is to analyze and categorize the primary issue.

1. Analyze the root cause of the problem using the PRD, the code, and the error message.
2. If multiple issues exist, return only the most severe and primary one (return exactly one ProblemType).
3. Return the result in strict JSON format with the following structure:
{
    "ProblemType": {
        "MainCategory": "Main error category",
        "SubCategory": "Specific sub-category of the issue",
        "Reasoning": {
            "SymptomAnalysis": "Observed abnormal behavior (in Chinese)",
            "RootCause": "Attribution analysis combining PRD and code (in Chinese)",
            "ErrorMechanism": "Technical explanation of how the error occurs (in Chinese)"
        }
    }
}

Below is the data provided to you:
PRD:
{prd}

Generated Code:
{results}

Error Message:
{input_text}

Please ensure your response strictly follows the JSON format above.

The allowed values for MainCategory are limited to the following five options - read them carefully and choose the most appropriate one:
1. Logic Error: Logical errors such as assertion failures or failure to meet PRD requirements.
2. Syntax Error: Syntax issues such as unexpected tokens, indentation problems, etc.
3. Parameter Error: The function required by the PRD is present, but input/output parameters are incorrect or missing.
4. Incomplete Error: Some required functions are entirely missing as per the PRD. Make sure to distinguish between a truly missing function and one that exists but contains logic or syntax errors.
5. Others: Any other issues that do not fit the above categories.

You must carefully select the MainCategory to ensure accuracy.
Do not return any MainCategory that is not listed above.
Do not return an empty MainCategory.
\end{lstlisting}
\end{tcolorbox}
\caption{Prompt Template for Error Type Classification.}
\label{fig: prompt for error classification}
\end{figure*}

\begin{figure*}
\begin{tcolorbox}[title=Docstring Augmentation Prompt]
\lstset{
    basicstyle=\normalfont\sffamily\footnotesize,
    breaklines=true,
    frame=none,
    columns=fullflexible,
}
\begin{lstlisting}[breaklines=true, basicstyle=\ttfamily\small]
# Context: The following Python code is provided for reference. It includes functions, classes, and other elements that provide context for the function or class below. Additionally, any constants or variables defined outside functions/classes are considered as part of the context and should be explained if used.

# Full Code:
```python
{full_code}
```

# Code for {name}:
```python
{code_snippet}
```

# Docstring:
Please generate a concise and clear docstring for the above {name} based on the full code context. Ensure the docstring briefly explains the {name}'s purpose, parameters, return values, and any relevant dependencies or interactions with other parts of the code. If there are any constants or variables used within the {name}, explain their role and significance, including where they are defined and how they interact with the function or class.

For functions: describe the input parameters, expected output, and any important side effects in a few sentences. Also, explain any constants used inside the function (if applicable).

For class.methods: describe the input parameters, expected output, and any important side effects in a few sentences. Also, explain any constants used inside the function (if applicable).

For classes: describe the main attributes and methods, along with the general purpose of the class in a brief summary. Mention any constants used in the class and explain their purpose and how they interact with class methods and attributes. Keep the docstring focused, avoiding unnecessary details or repetition.

# Output format
Your response should strictly follow the format below, without any other text or comments.
\"\"\"
docstring
\"\"\"
\end{lstlisting}
\end{tcolorbox}
\caption{Docstring Augmentation Prompt in Task Generation.}
\label{fig: prompt for docstring augmentation}
\end{figure*}

\begin{figure*}
\begin{tcolorbox}[title=Categories for Classifying Packages]
\lstset{
    basicstyle=\normalfont\sffamily\footnotesize,
    breaklines=true,
    frame=none,
    columns=fullflexible,
}
\begin{lstlisting}[breaklines=true, basicstyle=\ttfamily\small]
You are a Python expert. Given the name of a PyPI package, classify it into ONE category from the list below based on its MOST central and primary purpose.

Categories:

1. Web & Network Automation
Packages that support automation of web browsing, API communication, and network protocols.
Criteria: Enables browser control, HTTP requests, network operations, or web server handling.

2. Data Processing & Integration
Packages that extract, parse, or convert structured/unstructured data formats.
Criteria: Handles parsing or converting text, JSON, YAML, XML, or dates.

3. Security & Access Control
Packages that focus on authentication, authorization, or access control mechanisms.
Criteria: Implements rules, policies, or authentication methods.

4. Command-Line & Developer Tools
Packages that assist in building CLI tools, test frameworks, or code quality analysis.
Criteria: Aimed at improving the development experience, command-line interfaces, or code quality.

5. Cloud & Data Storage
Packages interacting with cloud services, databases, or data storage solutions.
Criteria: Provides interfaces or tools to access, manage, or validate remote data or cloud resources.

6. Data Science & Visualization
Packages used for scientific computing, visualization, or statistical evaluation.
Criteria: Supports data analysis, visualization, or scientific research.

7. Others
Packages that do not clearly belong in the other categories or are too general/specialized.
Criteria: Doesn't strongly align with the definitions above or serves a unique/niche purpose.

Please output only the category number (only one category), no explanation unless asked. Choose the single best fit.

Package name: {package_name}

You must strictly follow the format below, only a number no other text:
1
\end{lstlisting}
\end{tcolorbox}
\caption{Prompt Template for Classifying Packages.}
\label{fig: prompt for package classification}
\end{figure*}

\begin{figure*}
\begin{tcolorbox}[title=Human Evaluation Guideline for PRD Quality]
\lstset{
    basicstyle=\normalfont\sffamily\footnotesize,
    breaklines=true,
    frame=none,
    columns=fullflexible,
}
\begin{lstlisting}[breaklines=true, basicstyle=\ttfamily\small]
Each docstring is evaluated independently along the following three dimensions, using a 0-5 scale (0 = very poor, 5 = excellent):

Clarity - How easy the docstring is to understand for a competent software engineer. Consider language clarity, readability, and absence of ambiguity.
Completeness - Whether the docstring provides all necessary information to understand the function's behavior. Consider whether inputs, outputs, parameters, and important logic are described.
Actionability - How effectively the docstring guides actual implementation. Consider whether a developer could use the docstring alone to reasonably implement the function.

Rating Scale:
5: Excellent - No issues; highly clear, complete, and actionable.
4: Good - Minor improvements possible.
3: Fair - Understandable but lacking in one area.
2: Poor - Vague or missing key information.
1: Very Poor - Hard to follow or largely unhelpful.
0: Unusable - Cannot inform implementation at all.

If the original docstring is missing or boilerplate-only, please rate accordingly. Docstrings are to be rated individually without direct comparison.
\end{lstlisting}
\end{tcolorbox}
\caption{Human Evaluation Guideline for PRD Quality.}
\label{fig: instruct for prd quality}
\end{figure*}


\end{document}